\documentclass[prd,twocolumn,superscriptaddress,floatfix,amsmath,amssymb,amsfonts,nofootinbib,longbibliography]{revtex4-2}

\usepackage{float} 
\usepackage{scalerel}
\usepackage[normalem]{ulem}
\usepackage[english]{babel}
\usepackage{graphicx}
\usepackage{dcolumn}
\usepackage{bm}
\usepackage{blindtext}
\usepackage{verbatim}
\usepackage{relsize}
\usepackage{mathrsfs}
\usepackage{musicography}
\usepackage{amsmath}
\usepackage{blindtext}
\usepackage{cancel}
\usepackage{physics}
\usepackage{epstopdf}
\usepackage{mathtools}
\usepackage{blindtext}
\usepackage{tensor}
\usepackage{color}
\usepackage[usenames,dvipsnames]{pstricks}
\usepackage{epsfig}
\usepackage{pst-grad} 
\usepackage{pst-plot} 
\usepackage{hyperref}
\usepackage{verbatim}
\usepackage{slashed}
\usepackage{dsfont}

\hypersetup{
    colorlinks=true,
    linkcolor=blue,
    filecolor=red,      
    urlcolor=cyan,
}

\newcommand{\mf}{\mathsf}

\newcommand{\ii}{\mathrm{i}}
\allowdisplaybreaks[1] 

\begin{document}

\title{Localizing quantum fields with time-dependent potentials}

\author{Boris Ragula}
\email{bragula@uwaterloo.ca}
\affiliation{Department of Applied Mathematics, University of Waterloo, Waterloo, Ontario, N2L 3G1, Canada}
\affiliation{Institute for Quantum Computing, University of Waterloo, Waterloo, Ontario, N2L 3G1, Canada}
\affiliation{Perimeter Institute for Theoretical Physics, Waterloo, Ontario, N2L 2Y5, Canada}

\author{Bruno de S. L. Torres}
\email{bruno.desouzaleaotorres@ugent.be}
\affiliation{Perimeter Institute for Theoretical Physics, Waterloo, Ontario, N2L 2Y5, Canada}
\affiliation{Institute for Quantum Computing, University of Waterloo, Waterloo, Ontario, N2L 3G1, Canada}
\affiliation{Department of Physics and Astronomy, University of Waterloo, Waterloo, Ontario, N2L 3G1, Canada}
\affiliation{Department of Physics and Astronomy, Ghent University, Krijgslaan, 281-S9, 9000 Gent, Belgium}

\author{Erik Schnetter}
\email{eschnetter@perimeterinstitute.ca}
\affiliation{Perimeter Institute for Theoretical Physics, Waterloo, Ontario, N2L 2Y5, Canada}
\affiliation{Department of Physics and Astronomy, University of Waterloo, Waterloo, Ontario, N2L 3G1, Canada}
\affiliation{Center for Computation \& Technology, Louisiana State University, Baton Rouge, Louisiana, USA}

\author{Eduardo Mart\'{i}n-Mart\'{i}nez}
\email{emartinmartinez@uwaterloo.ca}
\affiliation{Department of Applied Mathematics, University of Waterloo, Waterloo, Ontario, N2L 3G1, Canada}
\affiliation{Perimeter Institute for Theoretical Physics, Waterloo, Ontario, N2L 2Y5, Canada}
\affiliation{Institute for Quantum Computing, University of Waterloo, Waterloo, Ontario, N2L 3G1, Canada}
\affiliation{Department of Physics and Astronomy, University of Waterloo, Waterloo, Ontario, N2L 3G1, Canada}

\begin{abstract}

In this paper we study the effect of localizing quantum field degrees of freedom by dynamically growing cavity walls through a time-dependent potential. We use our results to show that it is possible to do this without introducing non-negligible mixedness in localized modes of the field. We discuss how this addresses the concerns, raised in previous literature, that the high degree of entanglement of regular states in QFT may hinder relativistic quantum information protocols that make use of localized relativistic probes. 
\end{abstract}

\maketitle

\section{Introduction}

The issue of localization in relativistic quantum physics is a subtle one~\cite{Haag, Balachandran2020}. It has long been known that relativistic quantum field theories (QFTs) do not possess a meaningful notion of ``position operator'' that is consistent with relativistic causality~\cite{NewtonWigner, Wightman1962, noCausality1, Hegerfeldt1980, RUIJSENAARS198133, Malament1996}, and there cannot be any finite-rank projective measurements on local subregions of a QFT~\cite{Sorkin, Redhead1995, dowker2011}. Local measurements and operations, however, are some of the basic requirements of quantum information protocols that are fully covariant and causal in a relativistic sense. The subtle nature of localization in QFT thus poses a problem when studying the intersection of quantum information and quantum field theory. 


A versatile tool when dealing with local operations and measurements in QFT is that of a \emph{particle detector}. In Relativistic Quantum Information (RQI), a particle detector consists of a physically reasonable, localized quantum system that can be used to probe QFTs in local regions of spacetime. This is a concept that has proven to be very useful for several endeavours in the interface between quantum theory and relativity, ranging from studies of the Unruh effect and Hawking radiation~\cite{Unruh1976, Sciama1977, DeWitt} to the formulation of a measurement theory for QFT~\cite{jose}, as well as quantum information protocols in relativistic settings using quantum fields as mediators~\cite{cliche2010, landulfo2016, petar2020,QuantumCollectCalling, NoStrongHuygens, EricksonTeleportation, Pozas-Kerstjens:2015, Pozas2016}, among many others. 

The quantum systems used as particle detectors are typically conceptualized as internally nonrelativistic. This is in large part due to the close parallel between particle detector models in RQI and physical setups where the systems playing the role of the probe are accurately described by nonrelativistic quantum mechanics. A paradigmatic example of this is found in the light-matter interaction, where the internal dynamics of the atom (which acts in this context as a probe for the electromagnetic field) is well-approximated by nonrelativistic physics in most of the regimes of interest in atomic physics and quantum optics~\cite{richard}. The nonrelativistic approximations used in these models are very well-justified for most practical applications. However, it has been noted that using internally nonrelativistic systems as probes can also result in violations of relativistic causality and covariance, in particular when the detector is not spatially pointlike~\cite{us2, PipoFTL}. This has led to a recently renewed interest in probe models that are themselves given by quantum fields, as a way of reconciling the particle detector framework with the fully local and causal framework of QFT~\cite{fewster1,fewster3,MariaDoreen2023,QFTPD, QFTPDPathIntegrals}. 

A natural way to connect probe systems described by quantum fields with the more traditional nonrelativistic particle models is to single out a subset of modes of the probe field, and treat those modes as effective particle detectors. This is not just a convenient simplification: any experiment is limited by their energy budget available to fully resolve the high-energy sectors of the QFTs describing the probes. This is a practical yet fundamental limitation of any physical experiment. In particular, to reproduce the localization features of particle detectors, one would like to reduce the field theory to a finite number of modes that are also localized in space. However, the feasibility of fully relativistic local field modes as probes for quantum information protocols has recently been the object of some debate~\cite{max,RuepReply}. The main objection comes from the fact that, at a fundamental level, the restriction of the state of a QFT to a subset of degrees of freedom on a finite subregion will inevitably result in a mixed state, as a rather general consequence of the Reeh-Schlieder theorem~\cite{ReehSchlieder, witten}.

In the context of RQI, this fundamental level of mixedness presents a potential problem for fully relativistic implementations of \emph{entanglement harvesting}~\cite{max}, which is a protocol to extract entanglement from quantum field degrees of freedom in two spacelike separated regions~\cite{Pozas-Kerstjens:2015, Pozas2016}. The problem stems from the fact any initial mixedness in the probe systems will generally hinder their ability to become entangled through coupling to a quantum field. More broadly, any quantum information protocol in RQI may be potentially impacted by such unavoidable mixedness, as many protocols often rely on one's ability to prepare initially pure states with very high fidelity.

In this work, we will investigate to what degree a localized field can be used to produce truly local modes in a highly pure state when the localization profile of the field is implemented dynamically, by growing a potential that acts as an effective cavity localizing the field modes. 
By localizing a free field through a time-dependent confining potential, we will show it is possible to control to what extent the localized modes become mixed when starting from the free field vacuum. In particular, we will see that in the adiabatic limit (which in practice is most of the regimes where physical detectors are implemented), the mixedness becomes negligible. 

This manuscript is organized as follows. In Sec. \ref{sec:localFields} we present the formalism that will be used for localizing the free field. Section~\ref{sec:wightmanfunction} we review basic features of the field's two-point function, which is the main quantity that we will use to extract information about the effect of a time-dependent potential on a quantum field. In Sec. \ref{sec:numerics} we discuss the numerical methods that were used in order to produce all simulations throughout the manuscript. In Sec. \ref{sec:results} we show how an external potential can effectively confine  the energy carried by an initially localized wavepacket of the field. Section~\ref{sec:modemixedness} then shows that the dynamical creation of the potential walls can be done in such a way that localized modes of the field end up in states that are as close to pure as one desires. Finally, the conclusions of our investigation can be found in \ref{sec:conclusions}. 

\textit{Conventions.} We work in Minkowski spacetime with a Cartesian coordinate system  $x^\mu = (t, x^1, \dots, x^n)\equiv (t, \bm{x})$, where $n$ is the number of spatial dimensions. In these coordinates, the Minkowski metric takes the conventional form $\eta_{\mu\nu} = \text{diag}(-1, +1, \dots, +1)$. When referring to abstract spacetime points, we use sans-serif font $\mf{x}$. We use natural units, with $\hbar = c = 1$.

\section{Localized quantum fields}\label{sec:localFields}
The main object of interest in this work will be a linear scalar field $\phi(\mf x)$ under the influence of some external potential $V(\mf x)$. The theory in $d = n+1$ spacetime dimensions\footnote{Our numerical methods and results in later sections will be special to the case of $(1+1)$-dimensional Minkowki space; however, all of the structure in this section can be stated in arbitrary dimensions, so we chose to keep the more general language for now.} is described by the action
\begin{equation}\label{eq:KGaction}
    S = -\dfrac{1}{2}\int\dd^d x\left(\eta^{\mu\nu}\partial_\mu\phi\,\partial_\nu\phi + 2V(\mf x)\phi^2\right),
\end{equation}
which leads to the equation of motion
\begin{equation}\label{eq:KGequation}
    [\partial_\mu\partial^\mu -2V(\mf x)]\phi(\mf x) = 0.
\end{equation}
A general solution to Eq.~\eqref{eq:KGequation} can be written as
\begin{equation}\label{eq:modesum}
    \phi(\mf x) = \sum_j (a_{j} u_{j}(\mf x) + a^*_{j} u^*_{j}(\mf x)),
\end{equation}
for some set $\{u_{j}(\mf x)\}$ comprising a  (complex) basis to the space of solutions to the equations of motion. This basis is chosen such that
\begin{align}\label{eq:orthgonalityKGinnerproduct}
    (u_{j}, u_{k})_{\text{K.G.}} =& -(u_{j}^\ast, u_{k}^\ast)_{\text{K.G.}} = \delta_{jk}, \nonumber\\
    (u_{j}, u_{k}^\ast)_{\text{K.G.}} =& (u_{j}^\ast, u_{k})_{\text{K.G.}} = 0,
\end{align}
where the Klein-Gordon inner product $(f_1, f_2)_{\text{K.G.}}$ is given by
\begin{equation}\label{eq:KGinnerproduct2}
    (f_1, f_2)_{\text{K.G.}} = \ii \int_{\Sigma_t}\dd^n x\left(f^\ast_1\partial_t f_2 - f_2\partial_t f^\ast_1\right),
\end{equation}
and $\Sigma_t$ is a codimension-1 surface defined by a constant value of the time coordinate $t$ on a given inertial reference frame.\footnote{By using Stokes' theorem, we see that the Klein-Gordon inner product between any two functions $f_1$ and $f_2$ that satisfy the equation of motion~\eqref{eq:KGequation} does not actually depend on $t$, so we can choose to always evaluate Eq.~\eqref{eq:KGinnerproduct2} on $t=0$ without loss of generality.} 

In order to upgrade the classical field theory described by the action~\eqref{eq:KGaction} to a quantum field theory, we promote the coefficients $a_j, a^\ast_j$ in Eq.~\eqref{eq:modesum} to operators $\hat{a}_j, \hat{a}^\dagger_j$, satisfying the commutation relations 
\begin{align}\label{eq:CCRcreationannihilation}
       \big[\hat{a}_{j}^{\phantom{\dagger}},\hat{a}^\dagger_{k}\big] &= \delta_{jk} \openone\nonumber,\\
        \big[\hat{a}_{j}^{\phantom{\dagger}},\hat{a}^{\phantom{\dagger}}_{k}\big] &= 0 ,\\
        \big[\hat{a}_{j}^{{\dagger}},\hat{a}^\dagger_{k}\big] &= 0.\nonumber
\end{align}
In this case, Eq.~\eqref{eq:modesum} for the field $\phi(\mf x)$ becomes an expression for an operator (or more precisely, an operator-valued distribution) $\hat{\phi}(\mf x)$,
\begin{equation}\label{eq:modesumquantum}
    \hat{\phi}(\mf x) = \sum_j (\hat{a}_{j} u_{j}(\mf x) + \hat{a}^\dagger_{j} u^*_{j}(\mf x)),
\end{equation}
which satisfies the equal-time canonical commutation relations
\begin{equation}
    [\hat{\phi}(t, \bm x), \hat{\pi}(t, \bm x')] = \ii \,\delta^{(n)}(\bm x - \bm x')
\end{equation}
where $\hat{\pi}(t, \bm x) \equiv \partial_t\hat{\phi}(t, \bm x)$ is the conjugate momentum to the field $\hat{\phi}$.

    Upon quantization, the operators $\hat{a}^\dagger_j$ and $\hat{a}_j$ in the mode decomposition~\eqref{eq:modesumquantum} are interpreted as creation and annihilation operators, associated to modes of the field whose spacetime dependence is governed by the function $u_j(\mf x)$. To complete the process of canonical quantization, we construct the Hilbert space of the quantum field theory by starting with a vacuum state $\ket{0}$ that is annihilated by all of the annihilation operators,
    \begin{equation}
        \hat{a}_j\ket{0} = 0 \,\,\,\forall j,
    \end{equation}
    and then defining the QFT Hilbert space as the span of all the states obtained via repeated actions of creation operators on $\ket{0}$.
    
   If we set $V(\mf x) = m^2/2$ for some constant $m\geq 0$, then what we have just described is nothing but the theory of a Klein-Gordon field of mass $m$ in flat spacetime. In this case, one can choose the complete set of basis functions $\{u_j(\mf x)\}$ to be composed of \emph{plane waves}. For a theory in $n$ spatial dimensions, these are labeled by an $n$-dimensional wave vector $\bm{k}$, and can be written as
   \begin{equation}\label{eq:planewavemodes}
    u_{\bm k} (\mf x) \propto e^{\ii  k_\mu  x^\mu},
\end{equation}
where  $k^\mu = (\omega_{\bm k}, \bm{k})$, and $\omega_{\bm k}$ is determined by the dispersion relation
\begin{equation}
    \omega_{\bm{k}} = \sqrt{\abs{\bm k}^2 + m^2}.
\end{equation}
In free space, the wave vector $\bm{k}$ labeling different mode functions is continuous; we thus have to replace the sum over $j$ in Eq.~\eqref{eq:modesum} by an integral over $\bm{k}$, and the Kronecker deltas in Eqs.~\eqref{eq:orthgonalityKGinnerproduct} and~\eqref{eq:CCRcreationannihilation} by Dirac deltas. In the example of a field in a cavity with Dirichlet boundary conditions,  on the other hand, the possible wave vectors $\bm{k}$ are discrete, and the notation used in Eqs.~\eqref{eq:modesum},~\eqref{eq:orthgonalityKGinnerproduct},~\eqref{eq:CCRcreationannihilation} and~\eqref{eq:modesumquantum} is appropriate. 

It is also possible to consider cases where $V(\mf x)$ has some nontrivial spacetime dependence. Evidently, this will have an impact on the shape of the set of mode functions $\{u_j(\mf x)\}$ used as a basis of solutions to the equation of motion~\eqref{eq:KGequation}, and also influence whether this set is parametrized by discrete or continuous labels. If the potential is static---i.e., if $V(\mf x) = V(\bm{x})$ is independent of the inertial time coordinate in some inertial frame---one can still choose a basis of solutions which separate as $e^{-\ii\omega_j t}f_j(\bm x)$, but the general lack of translational invariance will mean that $f_j(\bm x)$ will not be given by pure plane waves. 

The case of most interest to us will be when the external potential $V(\bm x)$ grows to infinity as the distance between $\bm{x}$ and some finite region of space increases, such that all of the spatial profiles $f_j(\bm{x})$ will be strongly localized in a region around the minima of the potential. When this happens, we say that $\phi(\mf x)$ is a \emph{localized quantum field}: every admissible field configuration will effectively be confined in a localized region of space, determined by the region of strong support of the basis functions $f_j(\bm x)$.  QFTs with this feature have recently been shown to be quite useful to the construction of fully relativistic versions of particle detector models (see, e.g., \cite{fewster1, fewster3, QFTPD, QFTPDPathIntegrals}).

Our goal in the rest of the paper will be to study the price one pays when  dynamically constructing  localized quantum field theories from a free field theory, both from the energetic and the entropic point of view.  We will do so by starting with a free Klein-Gordon field and then applying a time-dependent external potentials $V(\mf x)$ which grows for a finite period of time, until it stabilizes in a shape that mimics a confining cavity, modelling the construction of a realistically confined field. 

\section{The dynamics of the two-point  function}\label{sec:wightmanfunction}

In this review section we will focus our attention on the two-point correlator of the field.  We will review its definition, its relevance for our purposes, and its explicit form in a few relevant cases. Additionally, we lay out the time evolution equations for the two-point function as an initial value problem, which will be key to the numerical approach employed in this paper. 


\subsection{Generalities about the two-point function}

Given a quantum scalar field $\hat{\phi}(\mf x)$ in a state $\hat{\rho}$, the two-point function is given by 
\begin{equation}\label{eq:GenWightman}
    W(\mf x, \mf x') = \expval{\hat{\phi}(\mf x)\hat{\phi}(\mf x')}_{\hat{\rho}} = \Tr(\hat{\rho}\,\hat{\phi}(\mf x)\hat{\phi}(\mf x')),
\end{equation}
where $\mf x$ and $\mf x'$ are two arbitrary spacetime points. 

The two-point function plays a central role in many aspects of quantum field theory. If the state $\hat{\rho}$ is Gaussian with vanishing first moments---as is the case of the vacuum, as well as any squeezed thermal state of a theory with linear equations of motion---then all the information about the quantum state $\hat{\rho}$ is encoded in $W(\mf x, \mf x')$. This allows for a much more efficient description of the physics of Gaussian states through properties of the two-point function~\cite{Sorkin_2014, FromGreenFunctiontoQFT, Sorkin_2018}. 
Moreover, even if $\hat{\rho}$ is not Gaussian, the two-point function contains all the information needed to specify the energy-momentum tensor of a field theory with quadratic action such as~\eqref{eq:KGaction}~\cite{birrell_davies, Wald2}.
Finally, the two-point function fully determines the leading-order response of the system to the action of an external agent that is linearly coupled to the quantum field. This plays an important role in the linear response theory of the quantum field under the influence of external sources, and also characterizes the effect of the field on the state of localized probes that couple to it. The latter application is of particular relevance, for instance, in the context of particle detectors in Relativistic Quantum Information~\cite{DeWitt, Louko_2006, cliche2010, martin-martinez2015, landulfo2016, petar2020, ericksonCapacity}.

Given its importance, it is useful to have explicit forms for the two-point function in a variety of situations. If we take the field state to be given by $\hat{\rho} = \ket{0}\!\bra{0}$, where we remember that $\ket{0}$ was defined as the state that is annihilated by all the annihilation operators $\hat{a}_j$ in the field expansion~\eqref{eq:modesumquantum}, then we have
\begin{equation}\label{eq:vacWightman}
    W_0(\mf x, \mf x') = \sum_j\, u_{j}(\mf x)u^*_{j}(\mf x').
\end{equation}
For a scalar field in free space with $V(\mf x) = m^2/2 = \text{const.}$, we can use the mode functions in terms of plane waves as stated in Eq.~\eqref{eq:planewavemodes} to find
\begin{equation}
    W_0(\mf x,\mf x') = \int\dfrac{\dd^n\bm k}{(2\pi)^n 2\omega_{\bm k}} \,e^{\ii k_\mu(x-x')^\mu}.
\end{equation}

Another ubiquitous  example is that of a one-particle Fock state, which is a natural way to construct wavepackets of the field that are initially localized in space. These states are generally of the form 
\begin{equation}\label{eq:freeSingleState}
    \ket{\psi} =\int \dd \bm k\, f(\bm k) \hat{a}^\dagger_{\bm k}\ket{0},
\end{equation}
where 
\begin{equation}\label{eq:freeSingleNormalization}
    \int\dd\bm k\, |f(\bm k)|^2 = 1.
\end{equation}
For this state, the two-point function becomes 
\begin{equation}\label{eq:stateWightman}
    W_\psi(\mf x, \mf x') = h(\mf x,\mf x') + W_0(\mf x, \mf x'),
\end{equation}
where 
\begin{equation}
    h(\mf x ,\mf x') = F(\mf x)F^*(\mf x') + F^*(\mf x)F(\mf x')
\end{equation}
is the state-dependent component of the two-point function, and
\begin{equation}
    F(\mf x) = \int\dd \bm k\, f(\bm k)u_{\bm k}(\mf x).
\end{equation}
More generally, it is possible to show that for any state $\ket{\psi}$ in the QFT Hilbert space,
the two-point function will be given by an expression like Eq. \eqref{eq:stateWightman}, where $h(\mf x, \mf x')$ will have a different form depending on the choice of $\ket{\psi}$. For details, see e.g. Appendix B of \cite{Talesgeometry}.

\subsection{Example in a cavity}
Our goal in this paper is to study the effects of a confining potential on the dynamics of an otherwise free quantum field. In order to prepare for that, it is instructive to review some basic features of the two-point function in a cavity. This will be useful because a cavity is nothing more than an idealized version of a confining potential that perfectly traps the quantum field in a finite spatial region. Moreover, as we will see in more detail in Sec. \ref{sec:numerics}, our numerical methods will inevitably rely on calculations in a finite computational domain, which requires us to specify how we treat the system's degrees of freedom at the boundaries of the domain. This ultimately acts as setting up an effective cavity in its own right. In this case, the cavity two-point function becomes the starting point for the field theory in free space, within the limits of our computational domain. In other words, what we call free space will be a large enough cavity, within which we will grow a smaller potential in a smaller region representing the realistic cavity. 

In this Section and for the rest of this paper, we will restrict to the case of 
a scalar field in $(1+1)$-dimensional spacetime. The field is then defined to be restricted within the region of space given by $0\leq x\leq L$, where $L$ is the total length of the cavity that constitutes our computational domain. For convenience, we choose to work with Dirichlet boundary conditions, for which \hbox{$W((0,t),(x',t')) = W((L,t),(x',t')) = 0$} and $W((x,t),(0,t')) = W((x,t),(L,t')) = 0$. This forces the wave numbers $\bm{k}$ labeling each mode function to become discrete, and given by $k_n = \frac{n\pi}{L}$, where $n \in \mathbb{N}$. We emphasize, however, that we will be mainly interested in features of the field in distance scales much smaller than $L$, which is chosen to be large enough so that we can show that our results will not depend on the boundary condition. Therefore, the choice of Dirichlet boundary condition for the purposes of our simulations should not limit the applicability of our results to the cases of interest in free space. 

In this particular setting, the basis modes for the field can be written as 
\begin{equation}
    u_{k_n}(\mf x) = \frac{1}{\sqrt{n\pi}}\sin(k_n x)e^{-\ii k_n t}.
\end{equation}
The vacuum two-point correlator then becomes 
\begin{equation}\label{eq:CavCorrelator}
    W_0(\mf x,\mf x') = \frac{1}{\pi}\sum_{n=1}^{\infty}\frac{1}{n}\sin(k_n x)\sin(k_nx')e^{-\ii k_n(t-t')}.
\end{equation}

By following a similar approach to the previous section, and making the appropriate substitutions to account for the discreteness of the modes in the cavity, we may construct a Fock wavepacket as 
\begin{equation}\label{eq:cavSingleState}
    \ket{\psi} = \sum_{n=1}^{\infty}f(n)\hat{a}^\dagger_{k_n}\ket{0},
\end{equation}
where $f(n)$ must be chosen such that 
\begin{equation}\label{eq:cavSingleNormalization}
    \sum_{n=1}^{\infty}|f(n)|^2 = 1.
\end{equation}
Here we can clearly see that Eqs. \eqref{eq:cavSingleState} and~\eqref{eq:cavSingleNormalization} are the cavity analogues to Eqs.~\eqref{eq:freeSingleState} and~\eqref{eq:freeSingleNormalization} respectively. Given this one-particle state, the resulting cavity two-point function is then 
\begin{equation}\label{eq:Cavity1Particle}
    W_\psi(\mf x,\mf x') = F^*(\mf x)F(\mf x') + F(\mf x)F^*(\mf x') +W_0(\mf x,\mf x'),
\end{equation}
where 
\begin{equation}
    F(\mf x) = \sum_{n=1}^{\infty}\frac{f(n)}{\sqrt{n\pi}} \sin(k_n x)e^{-\ii k_n t}.
\end{equation}

In both the continuum or the cavity case, we see that the state dependent component of the two-point function can be described through an integral transformation of the momentum profile for the state of interest. Additionally, the boundary conditions can introduce inhomogeneities in the spatial distribution of the modes. 
It is precisely the dynamics of Eq.~\eqref{eq:Cavity1Particle} that we will study with numerical methods. We are now ready to study the well-posedness of the evolution equations for the two-point function in a setting with boundary conditions.

\subsection{Initial value problem for the two-point function}

From the definition of the two-point function~\eqref{eq:GenWightman} and the fact that the field $\hat{\phi}(\mf x)$ satisfies the Klein-Gordon equation~\eqref{eq:KGequation}, it is easy to see that $W(\mf x, \mf x')$ will generally satisfy
\begin{align}
    \partial_t^2W(\mf x, \mf x') -\partial_x^2W(\mf x, \mf x') - 2V(\mf x) W(\mf x,\mf x')&=0\\
    \partial_{t'}^2W(\mf x, \mf x') -\partial_{x'}^2W(\mf x, \mf x') - 2V(\mf x') W(\mf x,\mf x')&=0.
\end{align}
That is, the correlation function satisfies the field equations of motion in both of its arguments. However, unlike the field amplitude $\hat{\phi}(\mf x)$ which is an operator-valued distribution, the two-point function $W(\mf x, \mf x')$ is a bi-scalar bi-distribution over spacetime. This makes it much easier to numerically evolve it using typical numerical methods for  partial differential equations (PDE).

We want to set up a well-posed boundary value problem for the two-point function (see, e.g., ~\cite{BenitoCauchyProblem}). In order to do that, we assume the field to be confined within a (large enough) domain $(x,x')\in [0,L]\cross[0,L]$, where we implement our boundary conditions at $x \text{ (resp. }x') = 0,L$. We thus obtain the full system, \begin{subequations}\label{eq:PDESys}
    \begin{equation}
        \partial_t^2W(\mf x, \mf x') -\partial_x^2W(\mf x, \mf x') - 2V(\mf x) W(\mf x,\mf x')=0,\\
    \end{equation}
    \begin{equation}
        \partial_{t'}^2W(\mf x, \mf x') -\partial_{x'}^2W(\mf x, \mf x') - 2V(\mf x') W(\mf x,\mf x')=0,\\
    \end{equation}
    \end{subequations}
    \begin{subequations}\label{eq:PDEICs}
    \begin{equation}
        W(\mf x,\mf x')\big |_{t = t' = 0} = W^{\phi\phi}(x,x'),\\
    \end{equation}
    \begin{equation}
        \partial_tW(\mf x,\mf x')\big |_{t = t' = 0} = W^{\Pi\phi}(x,x'),\\
    \end{equation}
    \begin{equation}
        \partial_{t'}W(\mf x,\mf x')\big |_{t = t' = 0} = W^{\phi\Pi}(x,x'),\\
    \end{equation}
    \begin{equation}
        \partial_t\partial_{t'}W(\mf x,\mf x')\big |_{t = t' = 0} = W^{\Pi\Pi}(x,x'),\\
    \end{equation}
    \end{subequations}
    \begin{subequations}\label{eq:PDEBCs}
    \begin{equation}
        W(0,t;x',t') = W(L,t;x',t') = 0,\\
    \end{equation}
    \begin{equation}
        W(x,t;0,t') = W(x,t;L,t') = 0. 
    \end{equation}
    \end{subequations}
The system described by Eqs.~\eqref{eq:PDESys},~\eqref{eq:PDEICs}, and~\eqref{eq:PDEBCs} is a hyperbolic PDE. It is therefore possible to show, under very general conditions, that this PDE system admits unique solutions for its advanced and retarded Green's functions. As shown in \cite{BenitoCauchyProblem}, if the initial data is prescribed for a two-point function, then the two-point function can be extended to the full spacetime through the causal propagator (retarded-minus-advanced) Green's function. This extension guarantees the well-posedness of the PDE system, for general (including possibly time-dependent) external potentials $V(\mf x)$. 

This result is of particular importance when studying numerical solutions to a PDE, as a well-posed PDE has a good chance of being numerical stable with standard methods. With this in mind, however, there are still particular restrictions that may need to be imposed on the numerical method to ensure the stability and convergence of a particular method. These restrictions will be discussed in detail in the next section. 


\section{Numerical methods}\label{sec:numerics}
In this section we present the numerical methods that we use to evolve the two-point function of the field. We will also analyze the stability and convergence of the methods to ensure that the results obtained through these numerical methods are reliable. 

\subsection{Regularization of short-distance singularities and choice of potential}
One notable issue with the vacuum correlator is that the sum in Eq. \eqref{eq:CavCorrelator} is singular in the coincidence limit $\mf x \to \mf x'$. In order to deal with that in our numerical methods, we consider a regularization of the vacuum correlator by replacing the pointlike arguments by evaluations over a thin smeared domain (in the same spirit as in, e.g.,~\cite{FirstBlurredDelta, CasalsBlurredDelta}). This is done by taking a spacetime smearing function $F_{\mf x}(\mf y)$ that corresponds to a smoothed out version of a Dirac delta distribution centered at the point $\mf x$, and then defining the regularized two-point function as 
\begin{equation}\label{SmearedW}
    \mathcal{W}(\mf x,\mf x') = \int \dd^2y\,\dd ^2 y'\, W(\mf y,\mf y') F_{\mf x}(\mf y)F_{\mf x'}(\mf y'),  
\end{equation}
where $\dd^2 y \equiv \dd\tau\, \dd y$ is the volume element in $(1+1)$-dimensional Minkowski spacetime with Cartesian coordinates $(\tau, y)$, with $\tau$ and $y$ being timelike and spacelike coordinates, respectively.
By taking the spacetime smearing to be of the form $F_{\mf x}(\mf y) = f_x(y)f_t(\tau)$ with
\begin{align}
    f_x(y) &= \frac{1}{\sqrt{2\pi \sigma_x}}e^{-(x-y)^2/2\sigma_x^2},\\
    f_t(\tau) &= \frac{1}{\sqrt{2\pi \sigma_t}}e^{-(t-\tau)^2/2\sigma_t^2},
\end{align}
the resulting regularized two-point function is then given by 
\begin{align}\label{eq:SmearedWightman}
    \mathcal{W}(\mf x,\mf x') = \frac{1}{\pi}&\sum_{n=1}^{\infty}\frac{1}{n}e^{-k_n^2\sigma_x^2}e^{-k_n^2\sigma_t^2}\nonumber\\
    &\times\sin(k_nx)\sin(k_nx')e^{-\ii k_n(t-t')}. 
\end{align}

\begin{figure}[h!]
    \centering
    \includegraphics[width=8.6cm]{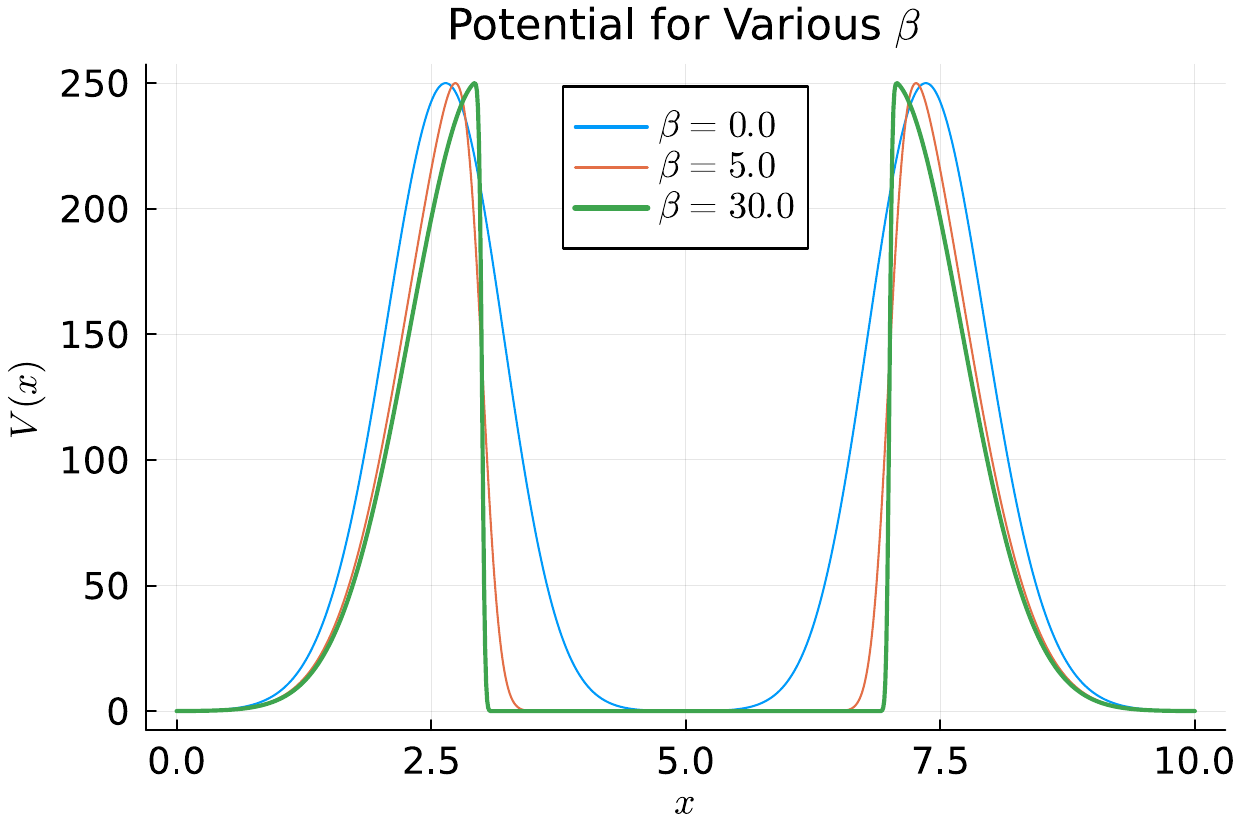}
    \caption{This plot shows the choice of confining potential for different choices of $\beta$ when $t\geq T$, with $V_{\text{max}} = 250$, $x_\textsc{l} = 3$, and $x_\textsc{r} = 7$. }
    \label{fig:Potentialvbeta}
\end{figure}

To complete the setup for our simulation, we have to pick the spatial and temporal profile of the external potential $V(x, t)$. We would like to choose the potential in such a way as to mimic a realistic confinement in a cavity; in particular, we consider the spatial profile of the potential to be a smooth function that  approximates a square cavity. As for the time dependence of the potential, we choose it to grow linearly in time, up to some predetermined time $T$. For concreteness, our potential is thus given by  
\begin{align}\label{eq:ThePotential}
    V(x,t)=
      V_{\text{max}}\frac{t}{T}&\bigg(e^{-(x-x_\textsc{l})^2/\ell^2}\left[1+\text{erf}(-\beta(x-x_\textsc{l}))\right] \nonumber\\
      &\!\!\!\!+ e^{-(x-x_\textsc{r})^2/\ell^2}\left[1+\text{erf}(\beta(x-x_\textsc{r}))\right]\bigg).
\end{align}
In this equation, $V_{\text{max}}$ is a prefactor chosen to control the height of the potential after it stops growing, $\beta$ determines how sharp the walls of the potential are, and $\ell$ determines the width of the potential walls. $x_\textsc{l}$ and $x_\textsc{r}$ are the location of the maxima of the left and right walls, respectively. In Fig. \ref{fig:Potentialvbeta} we show the potential after the cutoff time, $t>T$. From this figure, it is clear how this confining potential mimics that of a rectangular well. 
The efficacy of this potential will be explored in more detail in Sec. \ref{sec:results}.

\subsection{Numerical evolution of the two-point function}

In order to simulate the dynamics of the two-point function numerically, we use a second-order finite difference (FD) scheme in both time and space. The method works as follows. We begin by discretizing the domain of interest into a set of grid points, which we label $\{x_i\}_{i = 1}^{N_x}$ with $N_x$ being the number of grid points, with a uniform grid spacing of $\Delta x = x_i-x_{i-1}$. Similarly, for the temporal domain, we define $\{t_n\}_{n=1}^{N_t}$, and $\Delta t = t_{n} - t_{n-1}$. These grid points and grid spacing are adopted for both arguments of the two-point function so that both $(x,t)$ and $(x',t')$ have the same number of grid points in the spatial and temporal domains. We then define $\mathcal{W}_{ij}^{nm} = \mathcal{W}(x_i,t_n;x'_j,t'_m)$, where each label indicates the grid point on which the two-point correlator has been evaluated. For the confining potential, we similarly define $V^n_i = V(x_i,t_n)$. 

To now implement the equations of motion~\eqref{eq:PDESys} numerically, we will need to discretize the derivatives. For a function $f(x,t)$, we define the second order derivatives as
\begin{align}
    [D_t^2 f]_i^n &= \frac{1}{\Delta t^2}(f_i^{n+1}-2f_i^n+f_i^{n-1}),\\
    [D_x^2 f]_i^n &= \frac{1}{\Delta x^2}(f_{i+1}^n-2f_{i}^n+f_{i-1}^n),
\end{align}
where $[D_t^2 f]_i^n$ and $[D_x^2 f]_i^n$ denote the second derivative in time and second order derivative in space at the grid point $(x_i,t_n)$, respectively. Then, in order to evolve the two-point correlator, we will first fix a point $(\Tilde{x}',\Tilde{t}')$, and use our FD operators to evolve the two-point correlator using Eq. \eqref{eq:PDESys} in the form 
\begin{align}\label{eq:DiscretePDE}
    &\frac{1}{\Delta t^2}(\mathcal{W}_{ij}^{(n+1)m} -2\mathcal{W}_{ij}^{nm}+\mathcal{W}_{ij}^{(n-1)m}) = \nonumber\\*
    &\qquad\frac{1}{\Delta x^2}(\mathcal{W}_{(i+1)j}^{nm}-2\mathcal{W}_{ij}^{nm}+\mathcal{W}_{(i-1)j}^{nm})-2\Delta t^2 V^{n}_i\mathcal{W}^{nm}_{ij}.
\end{align}
Additionally, we have four initial conditions, and four boundary conditions. In this case, we adopt the following FD operators
\begin{align}
    [D_t\mathcal{W}]_{ij}^{00} &= \frac{1}{2\Delta t}(\mathcal{W}^{10}_{ij}-\mathcal{W}^{-10}_{ij}) = [\mathcal{W}^{\Pi\phi}]_{ij}^{00},\nonumber\\*
    [D_{t'}\mathcal{W}]_{ij}^{00} &= \frac{1}{2\Delta t}(\mathcal{W}^{01}_{ij}-\mathcal{W}^{0-1}_{ij}) = [\mathcal{W}^{\phi\Pi}]_{ij}^{00},\nonumber\\*
    [D_tD_{t'}\mathcal{W}]_{ij}^{00} &= \frac{1}{\Delta t^2}(\mathcal{W}^{11}_{ij}-\mathcal{W}^{01}_{ij}-\mathcal{W}^{10}_{ij}+\mathcal{W}^{00}_{ij}) \nonumber \\
    &= [\mathcal{W}^{\Pi\Pi}]_{ij}^{00}\label{eq:IC4}.
\end{align}
Note that, strictly speaking, our computational domain in time starts at $n, m = 0$; therefore, in the set of equations~\eqref{eq:IC4}, one should interpret the components $\mathcal{W}^{(-1)0}_{ij}$, $\mathcal{W}^{0(-1)}_{ij}$, and$\mathcal{W}^{11}_{ij}$ as being \emph{defined} by the conditions written above. This will be convenient when writing the recursive relation implementing the discrete time evolution in more concise form in what follows.

By solving for $\mathcal{W}^{n(m+1)}_{ij}$ in Eq. \eqref{eq:DiscretePDE}, we obtain the following recursive formula to evolve the two-point function on a constant $t$ slice: 
\begin{align}\label{eq:DiscreteEvolutiont'}
    \mathcal{W}^{n(m+1)}_{ij} = 2&\mathcal{W}_{ij}^{nm}-\mathcal{W}_{ij}^{n(m-1)}\nonumber\\
    &+\frac{\Delta t^2}{\Delta x^2}(\mathcal{W}_{i(j+1)}^{nm}-2\mathcal{W}_{ij}^{nm}+\mathcal{W}_{i(j-1)}^{nm})\nonumber\\
    &-2\Delta t^2 V^{n}_i\mathcal{W}^{nm}_{ij}.
\end{align}

Additionally, for the first time step on a constant $t$ slice, we can substitute $\mathcal{W}^{0(-1)}_{ij}$ from Eq.~\eqref{eq:IC4}. The form of this first time step is given by 
\begin{align}
    \mathcal{W}^{01}_{ij} = &\mathcal{W}^{00}_{ij}+\Delta t \mathcal{W}^{\phi\Pi}_{ij}\nonumber\\
    &+\frac{1}{2}C^2\left(\mathcal{W}^{00}_{i(j+1)}-2\mathcal{W}^{00}_{ij}+\mathcal{W}^{00}_{i(j-1)}\right)\label{eq:TimeStep1},
\end{align}
where $C = \frac{\Delta t}{\Delta x}$ is the Courant–Friedrichs–Lewy (CFL) factor. A similar approach can be taken for the $\mathcal{W}^{10}_{ij}$ term where we replace $\mathcal{W}^{\phi\Pi}_{ij}$ with $\mathcal{W}^{\Pi\phi}_{ij}$ and the second order spatial derivatives are taken with respect to the index $i$, and then solve for $\mathcal{W}^{11}_{ij}$ using the results of the evolutions for $\mathcal{W}^{01}_{ij}$ and $\mathcal{W}^{10}_{ij}$. Finally, in order to impose our initial condition $\mathcal{W}^{\Pi\Pi}(x,x')$, we simply use Eq. \eqref{eq:IC4}, along with result of the evolution in Eq. \eqref{eq:TimeStep1}. Using these particular equations involving our initial conditions for the first time step in each time direction gives us the necessary data to use our recursive timestepping in Eq. \eqref{eq:DiscreteEvolutiont'}. This will then generate the correlation function along the first two $t$-slices corresponding to $t_0$ and $t_1$.

Following this, we now have data for $\mathcal{W}^{0m}_{ij}$ and $\mathcal{W}^{1m}_{ij}$ for all $i,j,m$. Now we can evolve the two-point correlator forward along each $t$-slice using the following recursion relation
\begin{align}\label{eq:RecursionEvo}
    \mathcal{W}^{(n+1)m}_{ij} = 2&\mathcal{W}_{ij}^{nm}-\mathcal{W}_{ij}^{(n-1)m}\nonumber\\
    &+C^2(\mathcal{W}_{(i+1)j}^{nm}-2\mathcal{W}_{ij}^{nm}+\mathcal{W}_{(i-1)j}^{nm})\nonumber\\
    &-2\Delta t^2 V^{n}_i\mathcal{W}^{nm}_{ij}.
\end{align}

For the particular finite difference method that we use, it is important to ensure that the size of the time step $\Delta t$ is small enough for the solutions not to have spurious oscillations, which would render the method unstable~\cite{AtkinsonBook}. One common method of imposing bounds on the size of the time step is through von Neumann stability analysis~\cite{LeVequePDEs}. Applying a von Neumann stability test to our method, it can be shown (see Appendix \ref{AppendixBoris}) that the method is stable as long as
\begin{equation}\label{eq:CFLCondition}
    \Delta t^2 \leq \frac{\Delta x^2}{1+\frac{1}{2}\Tilde{V}\Delta x^2},
\end{equation}
where $\Tilde{V}$ is the maximum value attained by the potential. Therefore, in our simulations, we always choose a time step size in such a way that the condition \eqref{eq:CFLCondition} is satisfied. 
\subsection{Convergence tests}

When the exact solution to a set of differential equations with given initial conditions is known analytically, one can test the reliability/stability of a numerical method by simply comparing the solution generated by the numerical method with the benchmark analytic solution. However, even for a system of PDEs whose analytical solution is not known, one can still check that a numerical method produces reliable, convergent results in the following way. 

We first choose a grid spacing $\Delta x_c$ and $\Delta t_c$, where the subscript denotes that this is a \textit{coarse} grid spacing. We then define two more grid spacings $\Delta x_f = \frac{1}{h}\Delta x_m = \frac{1}{h^2}\Delta x_c$, where the subscripts $m$ and $f$ denote a \textit{medium} and \textit{fine} grid spacing relatively scaled by a factor $h\in\mathbb{Q}$, $h>1$. The pullback of the two-point function $\mathcal{W}$ to the coarse, medium, and fine grids then defines three sets of two-point correlators which we denote by $\mathcal{W}^{(c)}$, $\mathcal{W}^{(m)}$, and $\mathcal{W}^{(f)}$, respectively~\cite{LeVequePDEs}. Given each of the three two-point correlators associated to each of the grid spacings, we can check the spatial convergence order by plotting $\left(\mathcal{W}^{(c)}\right)^{\Tilde{n}\Tilde{m}}_{i\Tilde{j}} - \left(\mathcal{W}^{(m)}\right)^{\Tilde{n}\Tilde{m}}_{i\Tilde{j}}$ and $ h^p\left(\left(\mathcal{W}^{(m)}\right)^{\Tilde{n}\Tilde{m}}_{i\Tilde{j}} - \left(\mathcal{W}^{(f)}\right)^{\Tilde{n}\Tilde{m}}_{i\Tilde{j}}\right)$ for various points on the grid, where $\Tilde{n}$, $\Tilde{m}$, and $\Tilde{j}$ are fixed so that each solution is chosen at the grid points representing the same spacetime point in all three lattices, and the quantities $h$ and $p$ are two parameters characterizing the convergence test. When the method is convergent, we expect the two plots to coincide for suitably chosen values of those parameters.

The value of $p$ for which these two plots are expected to be convergent is known as the \emph{order of convergence}, and is determined by the order of the method used. The rescaling $h$, on the other hand, can be freely chosen by the user. In what follows, we will adopt the common choice of $h=2$, so that the scaling factor for the difference between the coarse and fine grids will be $h^2 = 4$. In our case of interest, where we use second-order finite difference schemes for our partial derivatives, the value of $p$ is known to be two. 

\begin{figure}[h!]
    \centering
    \includegraphics[width=8.6cm]{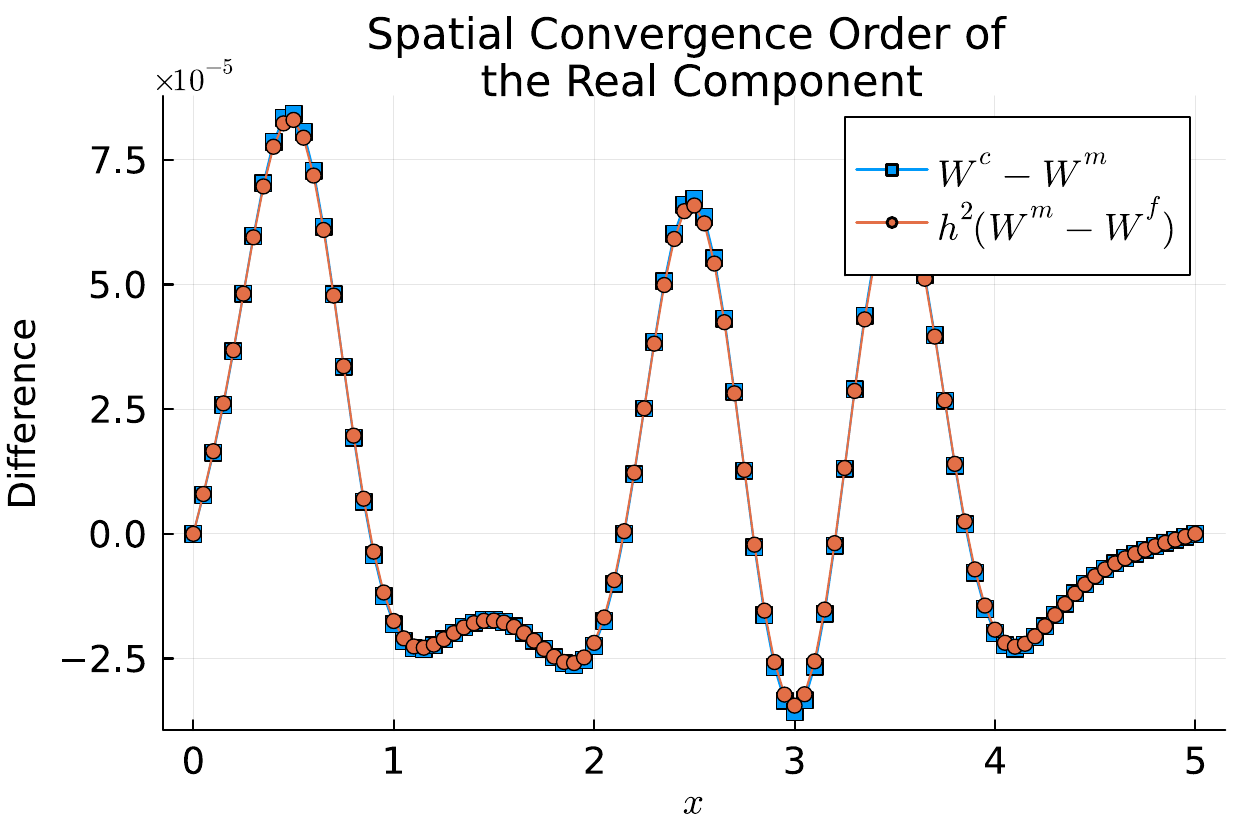}
    \caption{The spatial convergence order of the real component of the solution for the two-point function. For this convergence test, we chose $C = \frac{1}{5}$, with a spatial gridspacing of $\Delta x = 0.05$ with $N_x = 100$ for the coarse grid. We fixed the point $x' = 3.0$, $t = 0.75$ and $t' = 1.0$.}
    \label{fig:Real_Spatial_Convergence}
\end{figure}

\begin{figure}[h!]
    \centering
    \includegraphics[width=8.6cm]{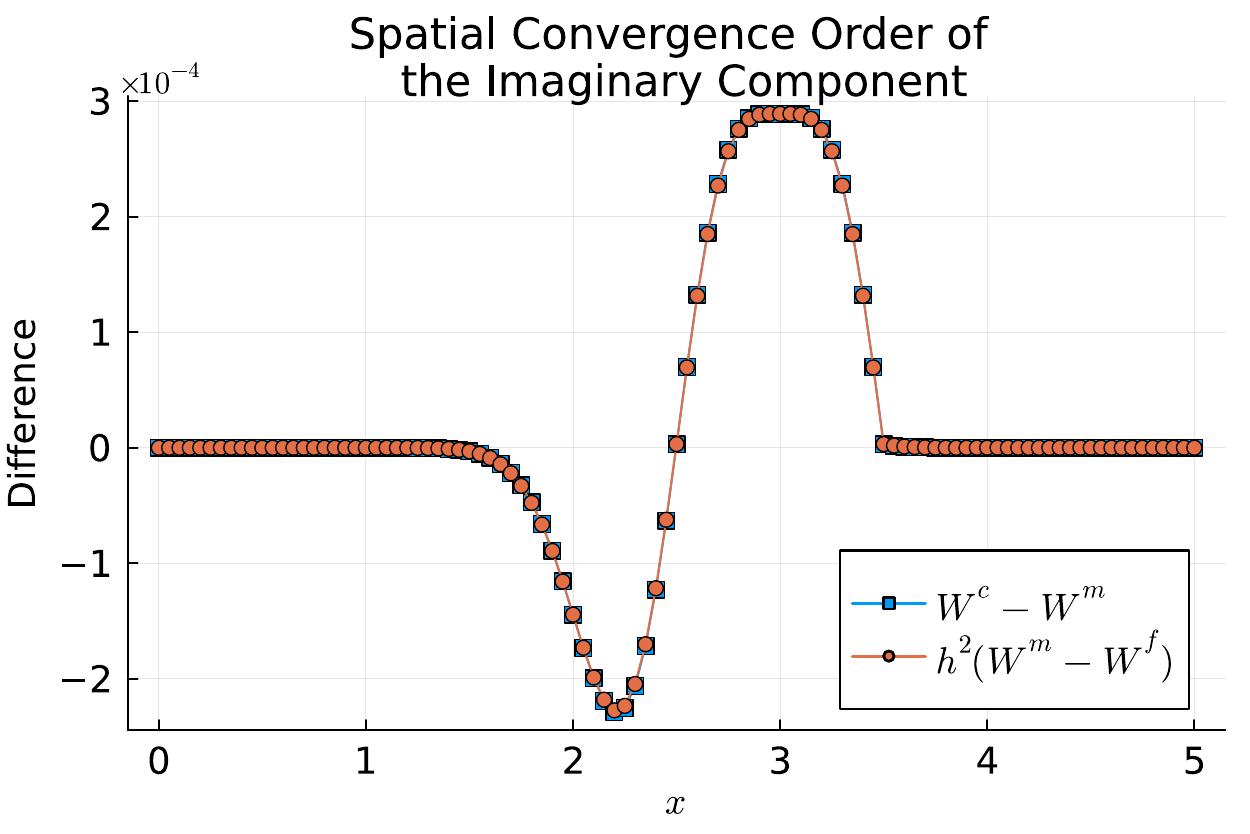}
    \caption{The spatial convergence order of the imaginary component of the solution for the two-point function. For this convergence test, we chose $C = \frac{1}{5}$, with a spatial gridspacing of $\Delta x = 0.025$ with $N_x = 100$ for the coarse grid. We fixed the point $x' = 3.0$, $t = 0.75$ and $t' = 1.0$.}
    \label{fig:Imag_Spatial_Convergence}
\end{figure}

The results of the spatial convergence test, where the quantities $\left(\mathcal{W}^{(c)}\right)^{\Tilde{n}\Tilde{m}}_{i\Tilde{j}} - \left(\mathcal{W}^{(m)}\right)^{\Tilde{n}\Tilde{m}}_{i\Tilde{j}}$ and $ h^p\left(\left(\mathcal{W}^{(m)}\right)^{\Tilde{n}\Tilde{m}}_{i\Tilde{j}} - \left(\mathcal{W}^{(f)}\right)^{\Tilde{n}\Tilde{m}}_{i\Tilde{j}}\right)$ are plotted at a constant time slice and one of the spatial coordinates $x$ is varied, are given in Figs.~\ref{fig:Real_Spatial_Convergence} and \ref{fig:Imag_Spatial_Convergence}, with parameters $p=h=2$. The spatial step was chosen to ensure that the solutions were resolved with enough accuracy to avoid any spurious errors, and the CFL factor was chosen to be within the bounds of Eq.~\eqref{eq:CFLCondition}. As we see, the two plots overlap almost exactly, confirming that the method is indeed second-order convergent in space. 

It is even more important to carefully assess the temporal convergence, which is more likely to be affected by numerical errors as the simulation runs. We can check the temporal convergence order by once again plotting $\left(\mathcal{W}^{(c)}\right)^{n\Tilde{m}}_{i\Tilde{j}} - \left(\mathcal{W}^{(m)}\right)^{n\Tilde{m}}_{i\Tilde{j}}$ and $ h^p\left(\left(\mathcal{W}^{(m)}\right)^{n\Tilde{m}}_{i\Tilde{j}} - \left(\mathcal{W}^{(f)}\right)^{n\Tilde{m}}_{i\Tilde{j}}\right)$ as before, with the only difference being that now we keep all coordinates fixed except for one of the timelike coordinates $t$.  The parameter $h$ can once again be chosen freely, whereas $p$ can be computed directly from the method. Just like before, we chose $h=2$, and note that the convergence order can be computed by~\cite{StrikwerdaBook,HildebrandBook} 
\begin{equation}\label{eq:timeconvergencetest}
    p(t) = \log_2\left(\frac{\norm{\left(\mathcal{W}^{(c)}\right)^{n\Tilde{m}}_{i\Tilde{j}} - \left(\mathcal{W}^{(m)}\right)^{n\Tilde{m}}_{i\Tilde{j}}}}{\norm{\left(\mathcal{W}^{(m)}\right)^{n\Tilde{m}}_{i\Tilde{j}} - \left(\mathcal{W}^{(f)}\right)^{n\Tilde{m}}_{i\Tilde{j}}}}\right).
\end{equation}
In this equation, we fix $\Tilde{m}$ and $\Tilde{j}$ and study the convergence only when one of the time coordinates is varying, as the symmetry of the two-point correlator implies that convergence along a $t$-slice is equivalent to convergence along the associated $t'$-slice. The norm $\norm{\cdot}$ in Eq.~\eqref{eq:timeconvergencetest} is chosen as the $L_2$-norm for the vector of gridpoint values at each time step. 
After the temporal convergence test is done, we find that $p(t)$ remains constant at $p(t)\approx 2.00$ for all values of $t$ considered in our paper. This confirms that the numerical method used is second-order convergent in time, as well as in space.

\section{Energy density behaviour in dynamical cavity creation}\label{sec:results}
\subsection{Energy-momentum tensor of the field}
Having established that our chosen numerical method is second-order convergent in both time and space, we can now move on to how we numerically obtain the energy density of the field from the evolution of the two-point correlator. In general, we can obtain the (expectation value of the) stress-energy tensor from the two-point correlator by evaluating
\begin{equation}\label{eq:vacuumsubtractedTuv}
    \langle\hat{T}_{\mu\nu}\rangle = \lim_{\mf x\rightarrow\mf x'}\Big(\left[P_{\mu\nu}-\eta_{\mu\nu}V(\mf x)\right] W_\textsc{r}(\mf x,\mf x')\Big),
\end{equation}
where $W_\textsc{r}(\mf x,\mf x') \equiv W(\mf x, \mf x') - W_0(\mf x,\mf x')$ is the renormalized two-point correlator (with $W_0(\mf x,\mf x')$ being the two-point correlator of the Minkowski vacuum), and
\begin{equation}
    P_{\mu\nu} \coloneqq \partial_\mu\partial'_{\nu}-\frac{1}{2}\eta_{\mu\nu}\partial_\alpha\partial'^{\alpha},
\end{equation}
where the symbol $\partial'_\nu$ stands for $\partial/\partial x'^\nu$---i.e., we take a partial derivative with respect to the $\nu$-th Cartesian coordinate of the second argument in $W_{\textsc{r}}(\mf x, \mf x')$.\footnote{Equation~\eqref{eq:vacuumsubtractedTuv} is a (highly simplified) version of the point-splitting regularization method, which allows one to compute expectation values of the energy-momentum tensor of a QFT in rather general settings, including curved spacetimes, by taking suitable limits of differential operators acting on two distinct spacetime points~\cite{Christensen1, Christensen2}.}
For the energy density, this reduces to
\begin{align}\label{eq:FieldEnergyDensity}
    \langle\hat{T}_{00}\rangle = \lim_{\mf x\to\mf x'}\frac{1}{2}\bigg[(\partial_t\partial'_{t}&+\partial_x\partial'_{x})W_{\textsc{r}}(\mf x,\mf x')\nonumber\\
    &+2V(x,t)W_{\textsc{r}}(\mf x,\mf x')\bigg].
\end{align}

To evaluate~\eqref{eq:FieldEnergyDensity} numerically, we can follow a similar approach to the one taken in Eq. \eqref{eq:IC4}, where we use a first derivative in both the $x,x'$ and the $t,t'$ coordinates to obtain
\begin{align}
    \langle\hat{T}_{00}\rangle_i^n = \frac{1}{2}\bigg([D_tD'_{t}\mathcal{W}_\textsc{r}]^{nm}_{ij}&+[D_xD'_{x}\mathcal{W}_\textsc{r}]^{nm}_{ij} \nonumber\\
    &+2V^n_i\mathcal{W}^{\phantom c nm}_{\textsc{r}\phantom c ij}
    \bigg)\bigg|_{n=m,i=j},
\end{align}
where $\mathcal{W}_\textsc{r}=\mathcal{W}-\mathcal{W}_0$  represents the smeared version of the renormalized two-point correlator as in Eq.~\eqref{SmearedW}. Once again, the CFL factor that we will use in our numerical calculations is $C=\frac{1}{20}$, which is  well within the bounds on the size of the time step for which~\eqref{eq:CFLCondition} is satisfied. We have checked that for this factor,  $p(t)\approx 2.00$ for all values of $t$ considered in our paper. Moreover, the spatial step was chosen to ensure that the solutions were resolved with enough accuracy to avoid any spurious errors.

\subsection{Testing the confining power of the effective cavity}

\begin{figure}[h!]
    \centering
    \includegraphics[width=8.6cm]{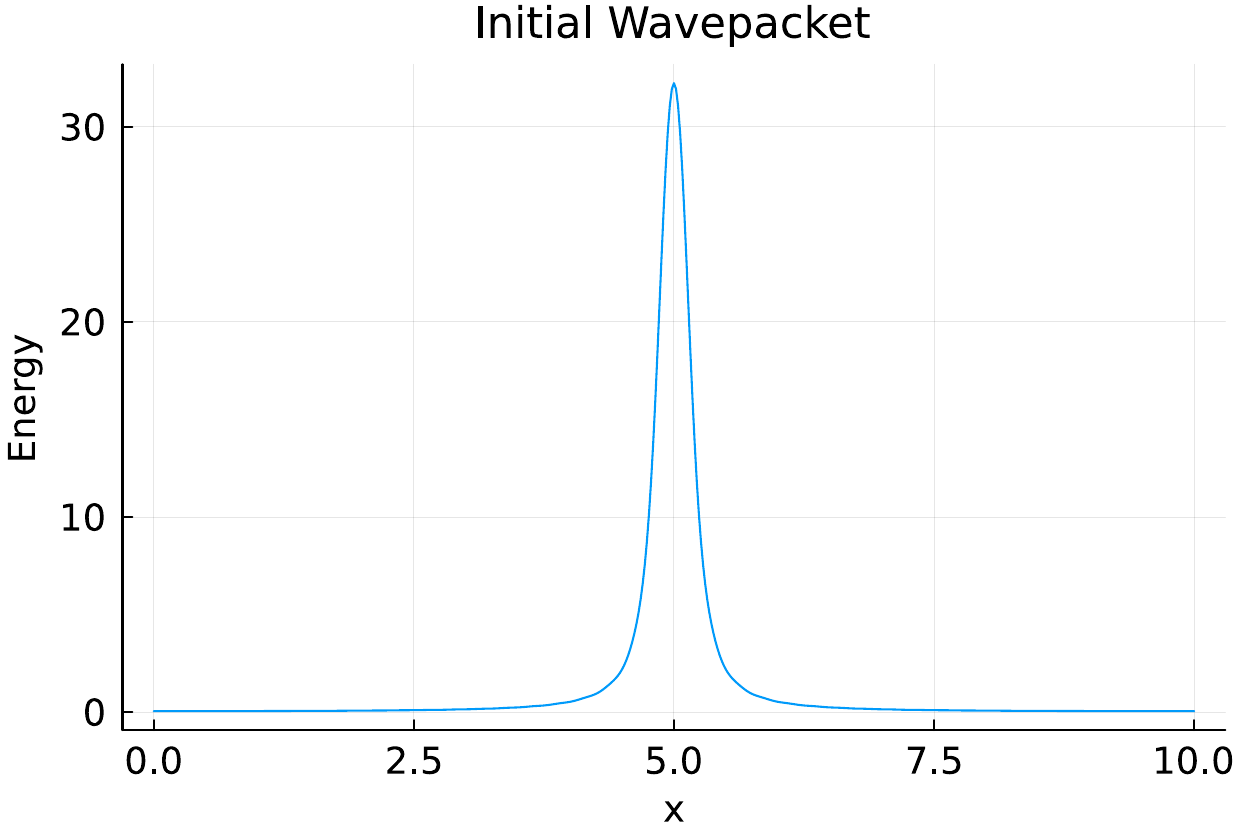}
    \caption{Initial energy distribution of the localized wavepacket, where $N = 100$ modes were used with $\alpha = 1/11$, and $x_0 = 5.0$.}
    \label{fig:Init_Wavepacket}
\end{figure}

Before doing a full dynamics simulation for the growth of the potential, it is instructive to test the extent to which the potential at late times (once the walls of the cavity are fully raised) is capable of  confining field excitations. In order to do so, we are going to test how good a static potential given by Eq.~\eqref{eq:ThePotential} at $t = T$ is at confining a localized wavepacket. We first define a state whose energy is localized within the interior region of the cavity. The particular wavepacket we chose is given by
\begin{equation}\label{eq:initialstatecavitytest}
    \ket{\psi} = \frac{1}{\lambda}\sum_{n=1}^N e^{-\alpha^2k_n^2/2}e^{\ii k_nx_0}\ket{n},
\end{equation}
where $\alpha$ is a parameter that controls the width of the wavepacket, $x_0$ is the spatial location of its peak, and $N$ is the number of modes we choose to include. $\lambda$ is just an overall normalization constant that ensures \mbox{$\braket{\psi}{\psi} = 1$}. The initial energy distribution of the field given this wavepacket can be visualized in Fig.~\ref{fig:Init_Wavepacket}.

\begin{figure}[h!]
    \centering
    \includegraphics[width=8.6cm]{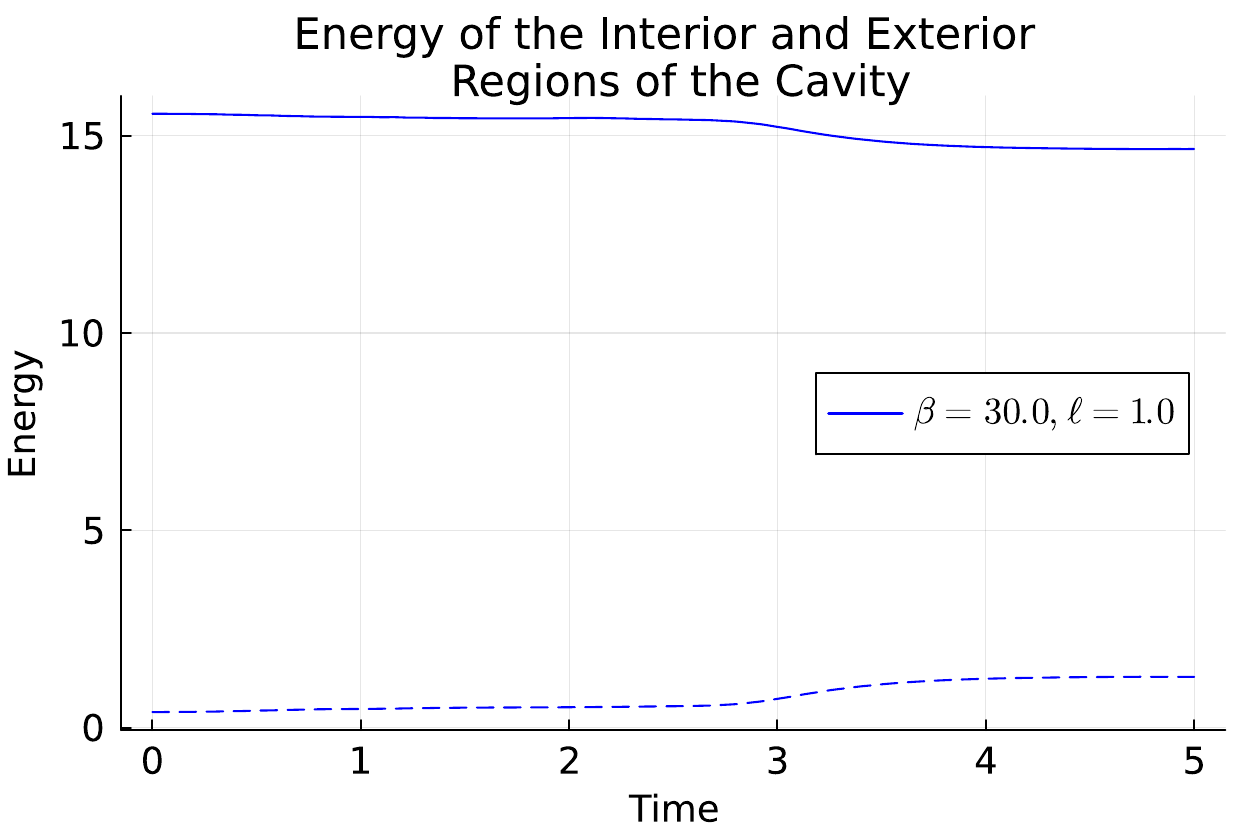}
    \caption{Energy as a function of time in the interior of the cavity and on the exterior of the cavity for $\ell = 1.0$, and $\beta = 30.0$. The cavity is chosen so that $x_\text{left} = 2.4$ and \hbox{$x_\text{right} = 7.6$}. The solid line represents the energy in the interior of the cavity, while the dashed line is the energy in the exterior region of the cavity.}
    \label{fig:EnergyByRegion}
\end{figure}

When comparing Fig. \ref{fig:Init_Wavepacket} to Fig. \ref{fig:Potentialvbeta}, we see that the field state is localized at the center of the confining potential with the peak located at $l = 2.5$ away from the edges. Then, we can plot the energy of the interior and exterior regions of the cavity as a function of time to determine how effective our cavity is at preventing transmission through the walls. The result can be seen in Figure~\ref{fig:EnergyByRegion}.

We see that the cavity is not perfectly confining as is the case for an infinite wall. This is expected, of course, since this potential is finite. In Fig.~\ref{fig:EnergyByRegion}, the left wall of the cavity is chosen to $x_\text{left} = 2.4$ and the right wall of the cavity is chosen to be $x_\text{right} = 7.6$. In comparing with Fig.~\ref{fig:Potentialvbeta}, we determined these values to be reasonably aligned with the outer boundaries of the potential.

At approximately $t = 2.75$, we see that some energy is able to penetrate the walls of the cavity. However, for the choice of parameters ($\beta=30$ and $\ell=1$), very little energy actually leaves the cavity, as illustrated by the overall small increase in energy in the exterior region. 

\begin{figure}[h!]
    \centering
    \includegraphics[width=8.6cm]{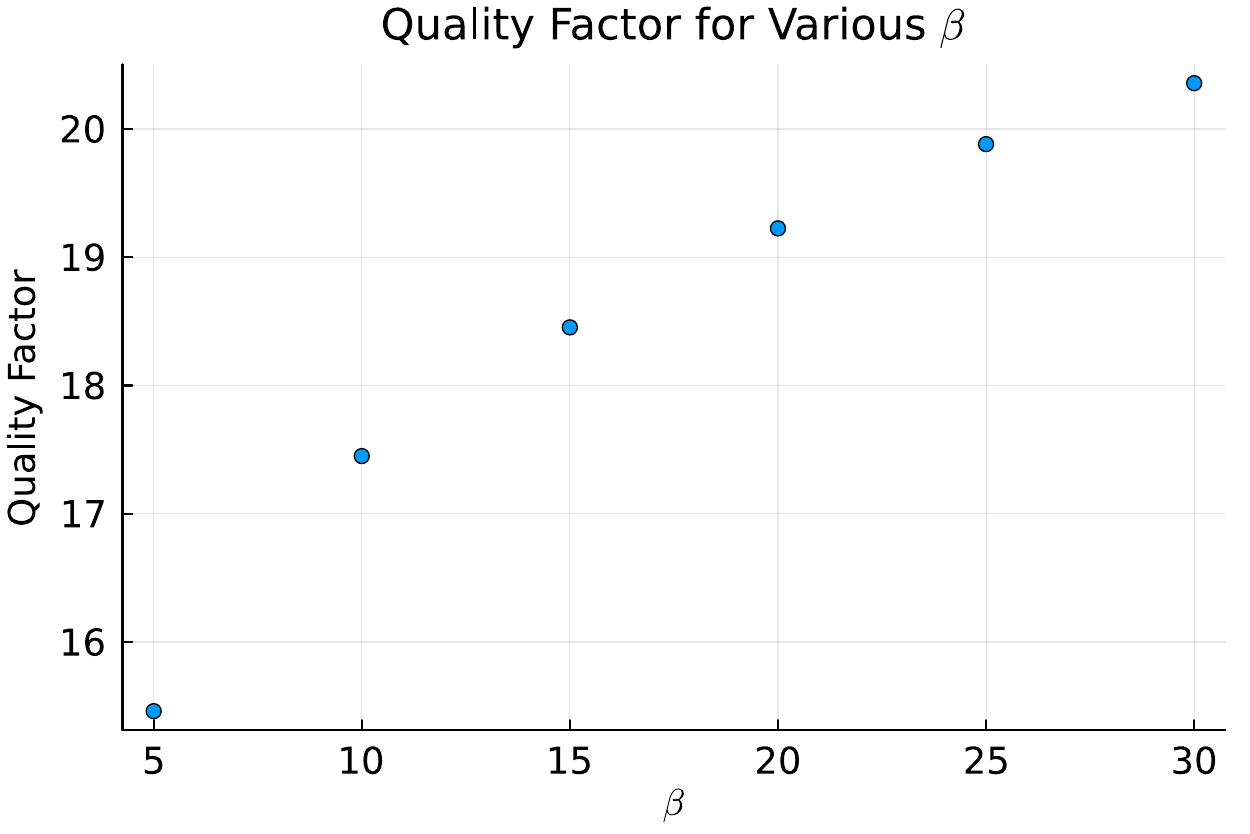}
    \caption{Quality factor of the confining potential as a function of $\beta$ (sharpness of the potential) for a fixed $V_{\text{max}} = 250$, $\ell = 1.0$, $x_\textsc{L} = 3.0$, and $x_\textsc{R} = 7.0$. We choose the boundaries of the cavity so that $x_\text{left} = 2.4$ and $x_\text{right} = 7.6$.}
    \label{fig:Qvbeta}
\end{figure}

\begin{figure}[h!]
    \centering
    \includegraphics[width=8.6cm]{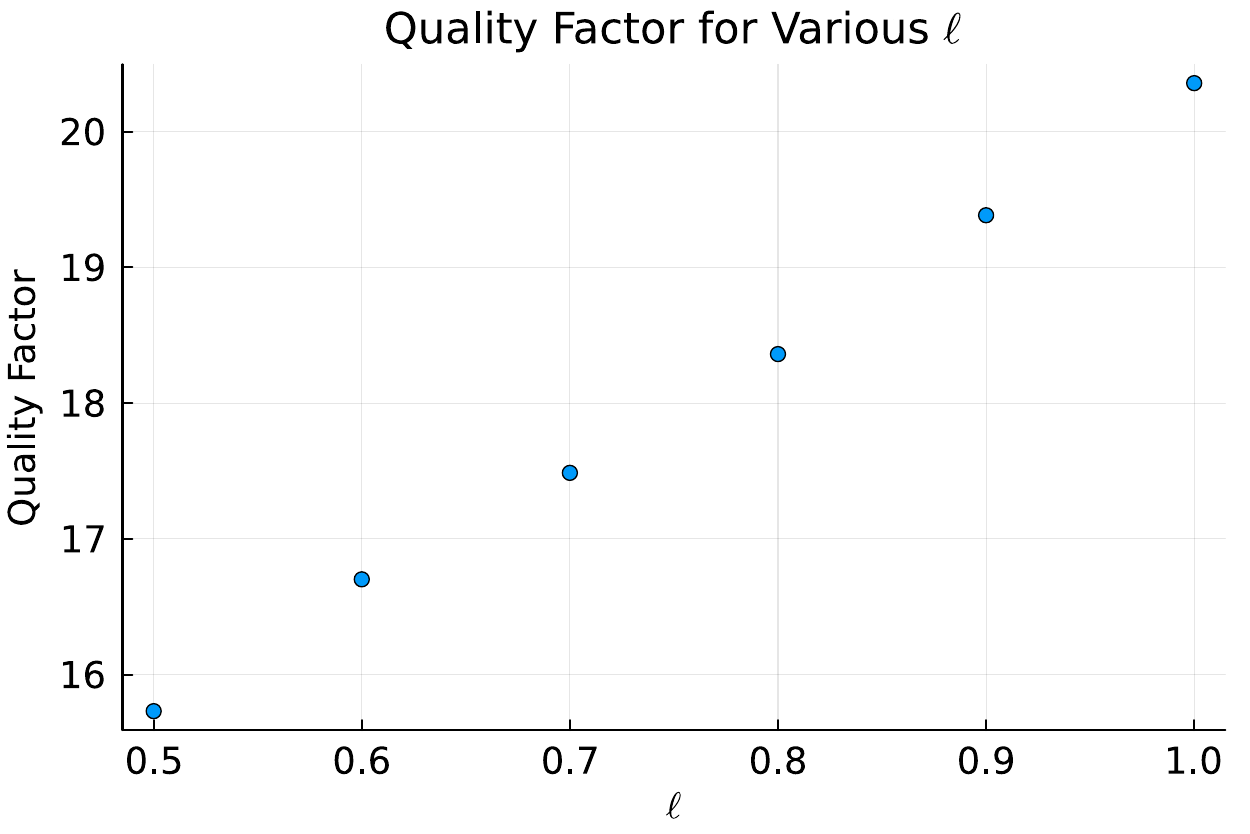}
    \caption{Quality factor of the confining potential as a function of $\ell$ (width of the wall) for a fixed $V_{\text{max}} = 250$, $\beta = 30.0$, $x_\textsc{L} = 3.0$, and $x_\textsc{R} = 7.0$. We choose the boundaries of the cavity so that $x_\text{left} = 2.4$ and $x_\text{right} = 7.6.$}
    \label{fig:Qvl}
\end{figure}

The potential's ability to partially confine the energy of the wavepacket between its peaks can also be visualized by looking at the quality factor $Q$ of the potential wall. The quality factor is defined as
\begin{equation}
    Q = \dfrac{(\text{initial energy stored inside at }t=0)}{(\text{energy lost in a cycle})},
\end{equation}
where, for a massless field in the setup of interest, ``one cycle'' of the system corresponds to one light-crossing time of the effective cavity whose walls are described by the external potential. In the case of our analysis, one-light crossing time is with respect to the cavity defined by $x_\text{left} = 2.4$ and $x_\text{right} = 7.6$, as discussed previously.
For a one-particle state of the form 
\begin{equation}
    \ket{\psi} = \sum_{n} c_n \ket{n},
\end{equation}
we can express $Q$ as
\begin{equation}
    Q = \dfrac{\sum_n \abs{c_n}^2 \omega_n}{\sum_n \abs{T(\omega_n) c_n}^2\omega_n},
\end{equation}
where $T(\omega_n)$ is the transmissivity of the potential for a mode with frequency $\omega_n$ that emanates from inside the cavity. For a general potential barrier, the transmissivity can only be computed numerically---and in fact, Figure~\ref{fig:EnergyByRegion} can be seen as an indirect numerical computation of the transmissivity of the effective cavity walls to the modes that contribute to the initial state~\eqref{eq:initialstatecavitytest}. The result for the quality factor of this effective cavity for various choices of parameters $\beta$ and $\ell$ can be found in Figures~\ref{fig:Qvbeta} and~\ref{fig:Qvl}.

In Fig. \ref{fig:Qvbeta}, we see that as we increase $\beta$, corresponding to the steepness of the interior wall of the cavity, the quality factor increases. In other words, the potential is better at retaining the modes that contribute to the initial state~\eqref{eq:initialstatecavitytest} when it grows more steeply in the region between its peaks.

In Fig. \ref{fig:Qvl}, we see that as the width of the potential increases, so does the quality factor. This is expected, of course, since we expect the transmissivity for typical modes within the cavity to leak through the potential barrier to decrease as the region where the potential is very large grows.

Overall, these results show that choosing a confining potential with $\beta = 30.0$ and $\ell = 1.0$ suffices in order to model an effective confining cavity for the short times at which we are investigating. We will thus use these values for the remaining analysis. 

\subsection{Dynamical creation of an effective cavity}

We now switch to the analysis of the dynamical creation of the cavity from empty space, with the field starting in the ground state. That is, we consider a scalar field theory with the time-dependent potential given by Eq.~\eqref{eq:ThePotential}, and take its initial state to be given by the vacuum of the free-space theory (that is, the massless scalar field without any external potential.) We will then analyze how the growth of the potential affects the energy density.

\begin{figure*}[t!]
    \centering
    \includegraphics[width = \textwidth]{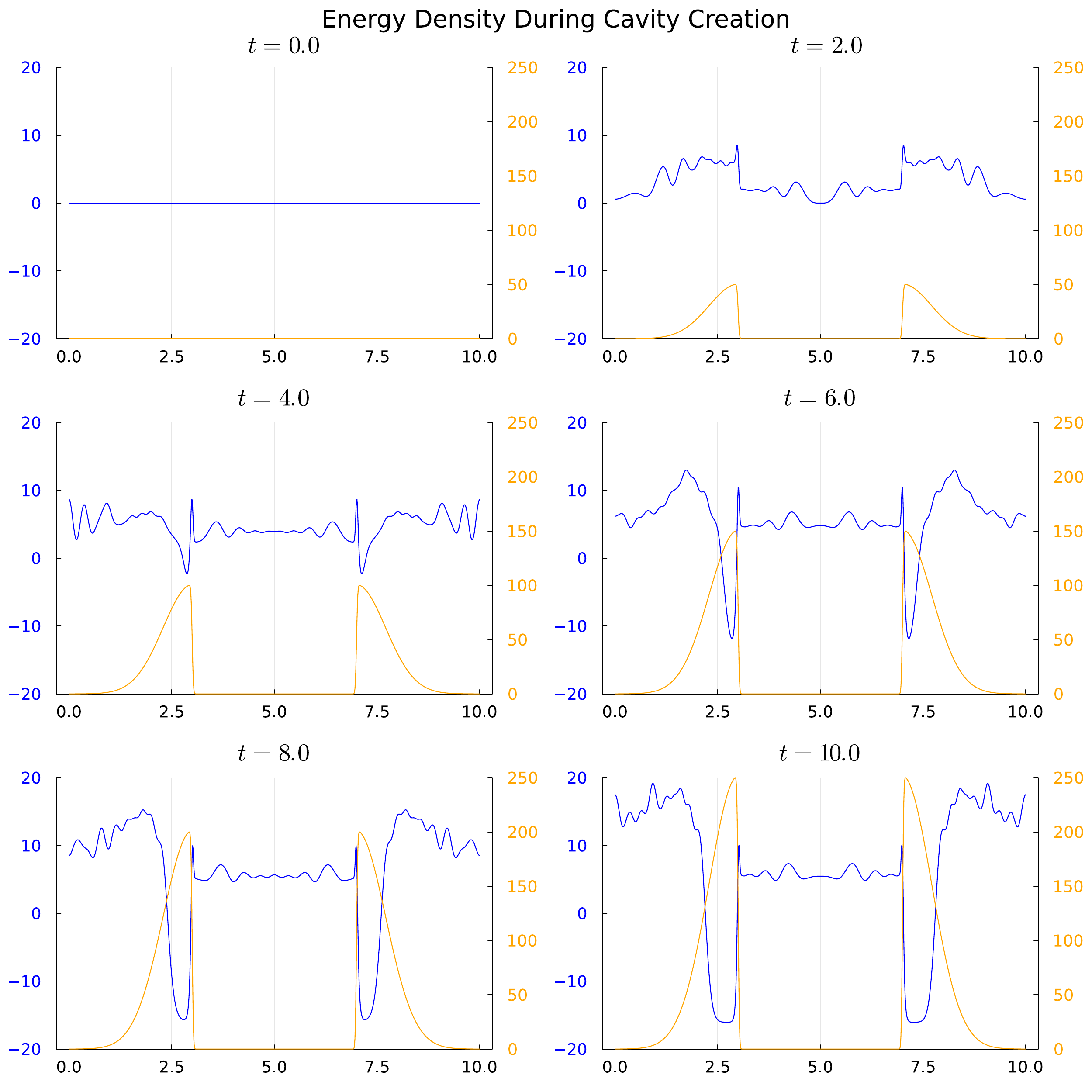}
    \caption{Energy density of the field during the time in which the cavity is being created. The blue curve is the energy density of the field whose value is given by the y-axis on the left of the plots, while the orange curve is the potential whose values are given by the y-axis on the right of the plots. For additional clarity, we have matched the y-axis colouring to that of the associated plot.}
    \label{fig:FinalT00}
\end{figure*}

\begin{figure}[h!]
    \centering
    \includegraphics[width=8.6cm]{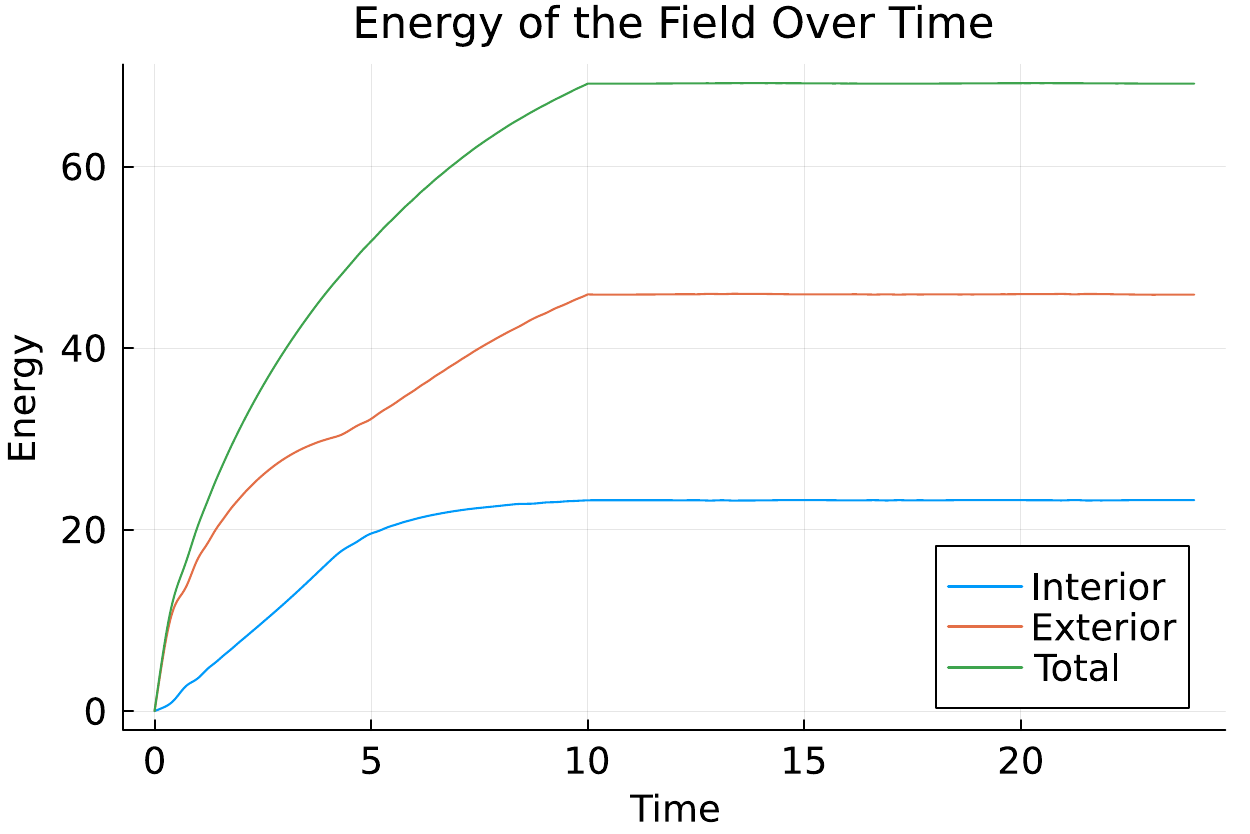}
    \caption{Energy of the field inside and outside of the cavity, as well as the total energy of the field as a function of time during a dynamic creation of the cavity. The potential grows up until $T = 10$ with $V_{\text{max}} = 250$.}
    \label{fig:RegionsOverTime}
\end{figure}

Since energy is injected into the field as the potential is created dynamically, one should in general expect the field to be excited---and therefore display a non-negligible energy density profile---at later times. The degree to which the field is excited, in turn, will depend on how fast the walls are created. In Fig. \ref{fig:FinalT00} we plot the energy density at different times during the creation of a potential that is grown until $T_\text{max} = 10$. 
In Fig. \ref{fig:RegionsOverTime} we again plot the energy in each particular region over time, where we see that after the potential stops growing, there is no change in energy on the interior region. This serves to confirm that the growth in the total energy in the field is indeed caused exclusively by the time dependence of the external potential, as expected---since we know that the total energy is a constant of motion in a system whose Hamiltonian is not explicitly time-dependent.

Notice that a faster growing potential injects more energy into the field during its growth. As a result, higher frequency modes of the cavity become excited. This will lead to more energy leaking through the potential barriers, as higher energy modes are more difficult to confine; conversely, the slower the growth of the potential, the better we expect it to be at confining excitations that are localized between its peaks.

Having established that the time-dependent potential acts as an effective cavity that confines the quantum field inside it, we can now finally turn our attention to the analysis of how the dynamical creation of the cavity affects the purity of the state localized and trapped inside.

\section{Effects of the dynamical potential on the mixedness of localized modes of the field}\label{sec:modemixedness}
Now that we have a way of dynamically localizing our free field by creating a cavity around it, the next question we investigate is that of the purity of the state of the degrees of freedom of the field trapped inside the cavity throughout the evolution. We will then later discuss why this is an interesting question to ask, especially from the point of view of the literature on Relativistic Quantum Information.

Before creating the potential, the vacuum state of the (free) field is a pure state; therefore, since the evolution of the field is unitary, the final global state of the field is also guaranteed to be pure. However, the act of creating the cavity walls can excite the field and introduce extra mixedness in the excitations in any local subregion of the field, and it is thus important to verify under what conditions these excitations can be controlled.

Intuitively, one would expect that it should be possible to arbitrarily decrease the deviation from purity by making the potential grow more slowly. This intuition follows from the adiabatic theorem, in which an adiabatically changing Hamiltonian will leave an initial ground state in the ``instantaneous'' ground state of the time dependent Hamiltonian for all times as long as the characteristic time scale over which the Hamiltonian is varying is much longer than the inverse of the energy gap between the ground state and the first excited state. We can test this intuition by quantitatively studying the loss of purity experienced by the field modes localized inside the created cavity (when the walls are raised in the presence of the vacuum state of the field) as a function of time in full detail. 

We start by first characterizing the degrees of freedom of the modes of the field that are localized inside the cavity. We do so by introducing the following localized quadrature operators: 
\begin{align}
    \hat{Q}(t) = \int\dd x f(x)\hat{\phi}(\mf x),\\
    \hat{P}(t) = \int\dd x\, g(x)\hat{\Pi}(\mf x),
\end{align}
where $\hat{\Pi}(\mf x) = \partial_t\hat{\phi}(\mf x)$ and $f(x)$ and $g(x)$ are the spatial profiles of the position and momentum normal modes of the cavity created by the confining potential. The functions $f(x)$ and $g(x)$ satisfy
\begin{equation}
    \int \dd x f(x)g(x) = 1,
\end{equation}
as a result of choosing $\hat{Q}$ and $\hat{P}$ such that $\comm{\hat{Q}}{\hat{P}} = \ii \mathds{1}$.

 Once the particular field mode and momentum mode profiles $f(x)$ and $g(x)$ are chosen, one can define the single-mode covariance matrix as 
\begin{equation}\label{eq:CovMatrix}
    \sigma = 
    \begin{pmatrix}
        2\langle\hat{Q}^2\rangle & 2\text{Re}\{\langle\hat{Q}\hat{P}\rangle\} \\
        2\text{Re}\{\langle\hat{Q}\hat{P}\rangle\} & 2\langle\hat{P}^2\rangle \\
    \end{pmatrix},
\end{equation}
where 
\begin{align}
    \langle\hat{Q}^2\rangle &= \int \dd x\dd x' f(x)f(x')W(\mf x,\mf x')\label{eq:Q2},\\
    \langle\hat{Q}\hat{P}\rangle &= \int \dd x\dd x' f(x)g(x') \partial_{t'}W(\mf x,\mf x')\label{eq:QP},\\
    \langle\hat{P}^2\rangle &= \int \dd x\dd x' g(x)g(x') \partial_t\partial_{t'}W(\mf x,\mf x')\label{eq:P2}.    
\end{align}
Given the covariance matrix $\sigma$ in Eq. \eqref{eq:CovMatrix}, we define the symplectic eigenvalue of $\sigma$ as the positive eigenvalue of $\ii \sigma\Omega^{-1}$, where 
\begin{equation}
    \Omega = 
    \begin{pmatrix}
        0 & 1 \\
        -1 & 0
    \end{pmatrix}
\end{equation}
is the symplectic matrix for a single bosonic mode. 
This leads to the following expression for the symplectic eigenvalue: 
\begin{equation}
    \nu(t) = 2\sqrt{\langle\hat{Q}^2(t)\rangle\langle\hat{P}^2(t)\rangle - \text{Re}\{\hat{Q}(t)\hat{P}(t)\}}.
\end{equation}

With that, the problem of computing the symplectic eigenvalue of the selected modes of the field boils down to obtaining the time evolution of the two-point function and evaluating the integrals~\eqref{eq:Q2}-\eqref{eq:P2}. We have already established that we can reliably evolve the two-point correlator numerically using the methods of Section~\ref{sec:numerics}. As for the integrations in Eqs. \eqref{eq:Q2}, \eqref{eq:QP}, \eqref{eq:P2}, these can be performed using a trapezoidal rule with the same grid spacing used to evolve the two-point correlator. The trapezoid rule for numerical integration for a function $f(x)$ is defined as 
\begin{equation}
    \int_a^b \dd xf(x) = \frac{\Delta x}{2}(f_0 + f_N) +\Delta x\sum_{i = 1}^{N-1}f_i,
\end{equation}
where $\Delta x$ is the grid spacing, and $N$ is the number of grid points that are used in the simulation. 

In order to extend this method to a $2$D integral over $x,x'$, we can use the following recursive method for multiple integration
\begin{equation}\label{eq:1DNumInt}
    \int_{a_x}^{b_x}\int_{a_y}^{b_y}\dd x\dd y f(x,y) = \int_{a_x}^{b_x}\dd x\, G(x),
\end{equation}
where 
\begin{equation}\label{eq:2DNumInt}
    G(x) = \int_{a_y}^{b_y}\dd y f(x,y).
\end{equation}
The result of evaluating~\eqref{eq:2DNumInt} using the prescription~\eqref{eq:1DNumInt} is to weight the ``corner'' gridpoints (i.e. $i=j=0$, $i=0,j=N_x$, $i=N_x, j = 0$, and $i=j=N_x$) by a factor of $\frac{1}{4}$, the ``edge" points (i.e $i\in \{1,N_x\}, j\in [2,N_x-1]$ and $i\in [2,N_x-1], j\in \{1,N_x\}$) by a factor of $\frac{1}{2}$ and all interior points (i.e $i,j\in [2,N_x-1]$) by a factor of 1. This method is well known to be second-order convergent.

Now, in order to evaluate the level of mixedness introduced in the modes of the field that remain trapped in the cavity due to its dynamical creation, we need to 1) pick a set of modes that represent physically accessible modes inside the cavity, and 2) analyze the state of these modes after the cavity is created after tracing out everything else.

\begin{figure}[h!]
    \centering
    \includegraphics[width=8.6cm]{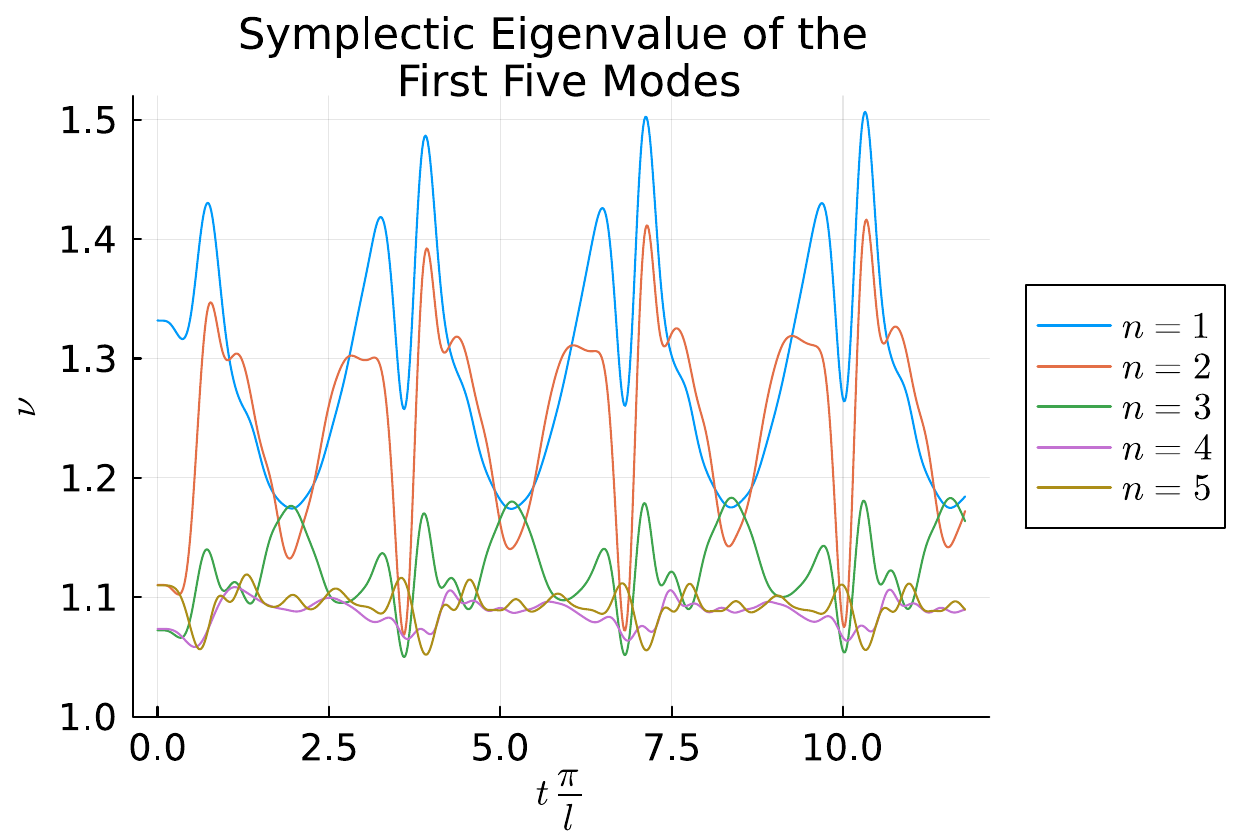}
    \caption{Symplectic eigenvalue for the first five modes of the field as a function of time (in units of the (inverse) fundamental frequency of the cavity). We raise the potential until $T = 10$ and with $V_{\text{max}} = 250$. The first mode gets more mixed because the switching is less adiabatic when the energy of the mode of study is lower.}
    \label{fig:SymplecticEigenvalues}
\end{figure}

Selecting a suitable profile for the quadratures requires some care. These profiles can be chosen, for example, by taking into account what are the actual degrees of freedom that are measurable by an experimenter; or alternatively, if we know the shape of the cavity, one can reasonably choose the normal modes associated with the eigenvectors of the d'Alembert operator inside the region of confinement. For our particular study, we notice from Fig.~\ref{fig:RegionsOverTime} that, once the potential has stopped growing, the energy within the cavity remains constant. Therefore, to a good approximation, we can say that the resulting cavity behaves as a (fully reflecting) Dirichlet cavity. As a result, we approximate the normal modes of the cavity by  the field amplitude and momentum spatial profiles as
\begin{equation}\label{eq:NormalMode}
    f(x)=g(x) = \sqrt{\frac{2}{l}}\sin(\frac{n\pi x}{l}),
\end{equation}
where $l$ is the length of the artificial cavity and $n$ corresponds to the mode number\footnote{Technically the momentum profile would be multiplied by an extra power of $l$, however choosing the same profile for amplitude and momentum can be achieved by a single-mode squeezing which does not affect the symplectic eigenvalue calculation for this mode.}. In Fig.~\ref{fig:SymplecticEigenvalues} we show the variation in time of  the symplectic eigenvalue of these effective normal modes as the cavity is created and how they behave after its creation. 

In Fig. \ref{fig:SymplecticEigenvalues} there are quite a few interesting features. The first is that initially our symplectic eigenvalue does not equal one. This is because any localized mode of a quantum field theory is guaranteed to be mixed, due to the inherent entanglement in the vacuum state; a small amount of mixedness will always appear as a consequence of the partition into modes, even before the cavity walls are raised. This simple fact has been used~\cite{max} to argue against the feasibility of such local field modes as probes for relativistic quantum information protocols such as entanglement harvesting; however, this conclusion is a bit too premature, as we will see shortly.

We also see from Fig. \ref{fig:SymplecticEigenvalues} that the purity fluctuates throughout the evolution, with its peaks in time occurring at intervals of one light crossing time of the artificial cavity. The fact that the oscillations have this periodicity is further confirmation that, within the timescales considered in this problem, the potential is truly behaving as a reflecting cavity. More quantitatively, the observed pattern of fluctuations is likely due to a combination of factors such as the fact that the modes of the cavity are not the exact normal modes for the local subregion, the fact that the potential is not infinitely confining, and the fact that the dynamical process of creating the cavity creates mixedness between the modes themselves and the outside degrees of freedom. Additionally, we see that the higher modes of the artificial cavity are less mixed than the lower modes. This is once again consistent with the intuition from adiabaticity arguments, since the adiabatic approximation is expected to be better if applied to modes of higher frequencies. In short, the adiabatic approximation becomes valid when $n \pi T/l\gg 1$.

In order to fully understand the effects of the speed at which the potential grows on the mixedness of the confined degrees of freedom of the field, we ran multiple simulations with varying values of $T$ to analyze the results on the symplectic eigenvalue of the modes associated to the spatial profile~\eqref{eq:NormalMode}. 

\begin{figure}[h!]
    \centering
    \includegraphics[width=8.6cm]{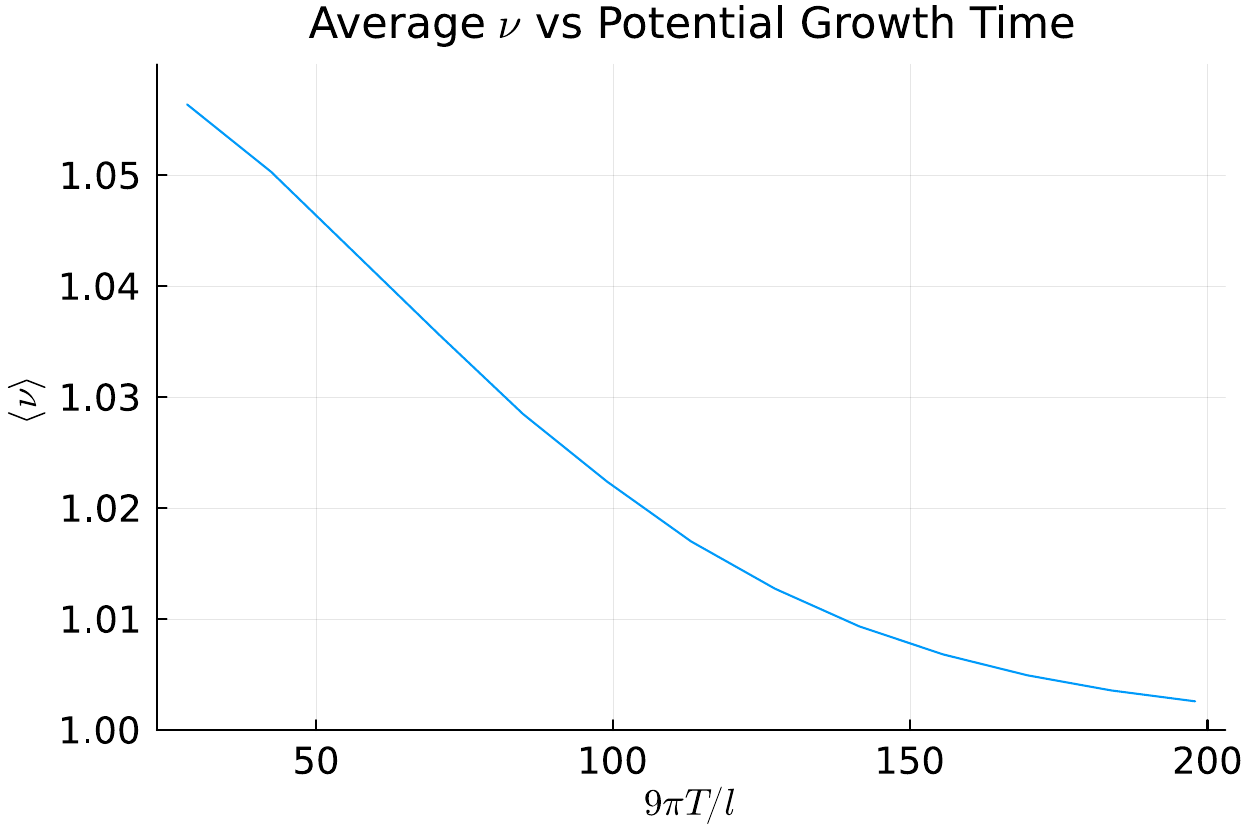}
    \caption{Average symplectic eigenvalue as a function of the stop time for the confining potential. Here we scale the x-axis by the frequency of the $n=9$ mode, akin to what was done for the previous plot of the symplectic eigenvalues}
    \label{fig:AverageSymplectic}
\end{figure}

Because the field will in general not be left in an exact eigenstate of the Hamiltonian for any finite-time simulation with the dynamical potential, the symplectic eigenvalue will oscillate as a function of time even after the potential stops growing. For better visualization, it is therefore more convenient to plot the time average of the symplectic eigenvalue as a function of $T$, where the average was taken over three times the light-crossing time of the effective cavity. This is what is displayed in Figure~\ref{fig:AverageSymplectic}. 

More specifically, Fig.~\ref{fig:AverageSymplectic} shows how the mixedness of the mode $n=9$ varies as a function of $T$. We choose $n=9$ as a computational compromise: this choice allows us to see the asymptotic behaviour of the purity easily within the confines of the computational resources that were available. The behaviour of any other mode is expected to be similar, only the timescale at which each mode will asymptote to near purity would vary proportionally to $1/n$. 

\begin{figure}[h!]
    \centering
    \includegraphics[width=8.6cm]{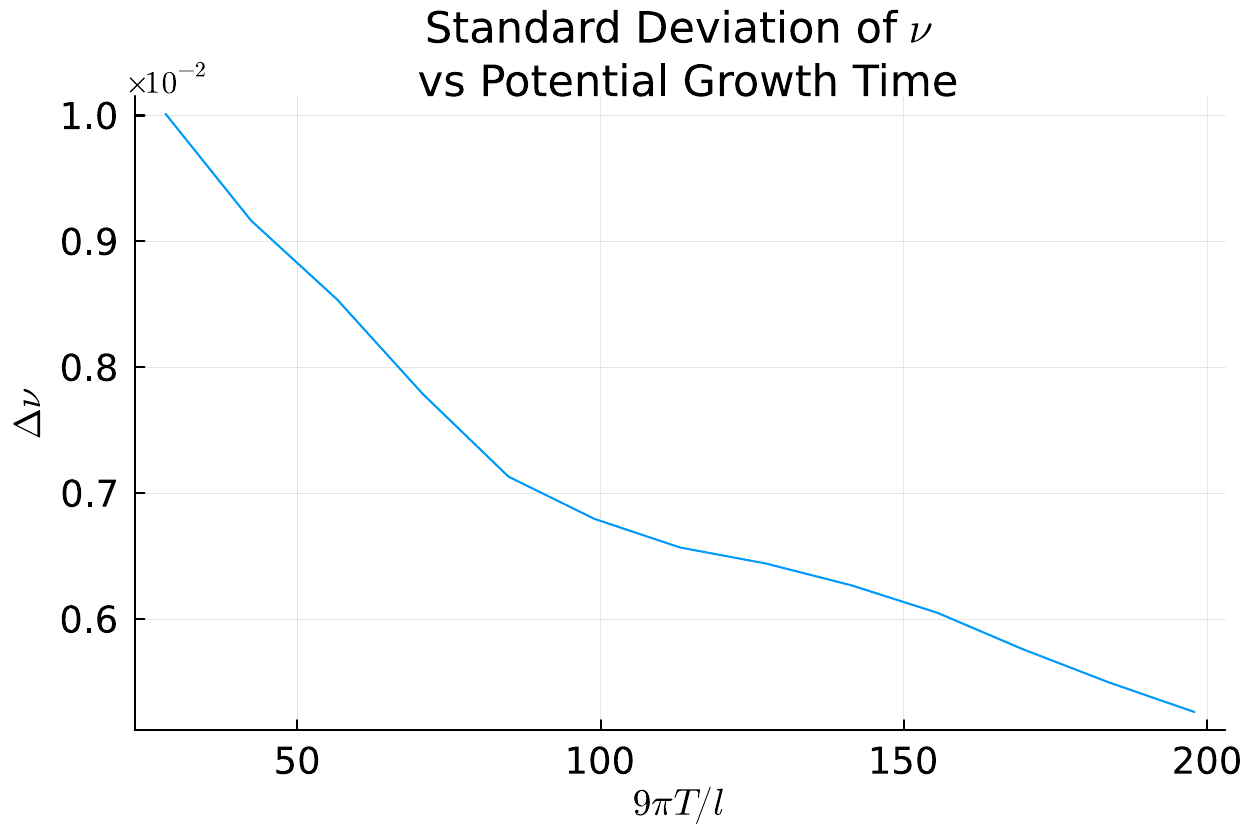}
    \caption{Standard deviation of the symplectic eigenvalue as a function of the stop time for the confining potential. Here we scale the x-axis by the frequency of the $n=9$ mode.}
    \label{fig:VarianceSymplectic}
\end{figure}

In this plot, we can see that as the creation of the cavity is more adiabatic, we see a monotonic decrease in the symplectic eigenvalue that tends towards purity \mbox{($\langle\nu\rangle=1$)}.\footnote{Here, in Figures~\ref{fig:AverageSymplectic} and~\ref{fig:VarianceSymplectic}, and for the rest of this section, the notation $\expval{A}$ will refer to the time average of the quantity $A$ over one light-crossing time of the effective cavity. This should not be confused with the expectation value of a given observable on a quantum state.} This corroborates the intuition from the adiabatic theorem that if you create the cavity slowly enough, the purity of the confined modes is not compromised. 

Given that the symplectic eigenvalues of the modes we choose are not constant over time, it is also important to make sure that its fluctuations do not go out of control. In order to quantitatively account for these fluctuations, we can study the standard deviation of the symplectic eigenvalue  $
    \Delta\nu = (\langle\nu^2\rangle-\langle\nu\rangle^2)^{1/2}$ as a function of $T$. We show the behaviour of the standard deviation in Fig.~\ref{fig:VarianceSymplectic}, where we see that it is small enough and also monotonically decreases as the cavity creation becomes more adiabatic.

\begin{figure}[h!]
    \centering
    \includegraphics[width=8.6cm]{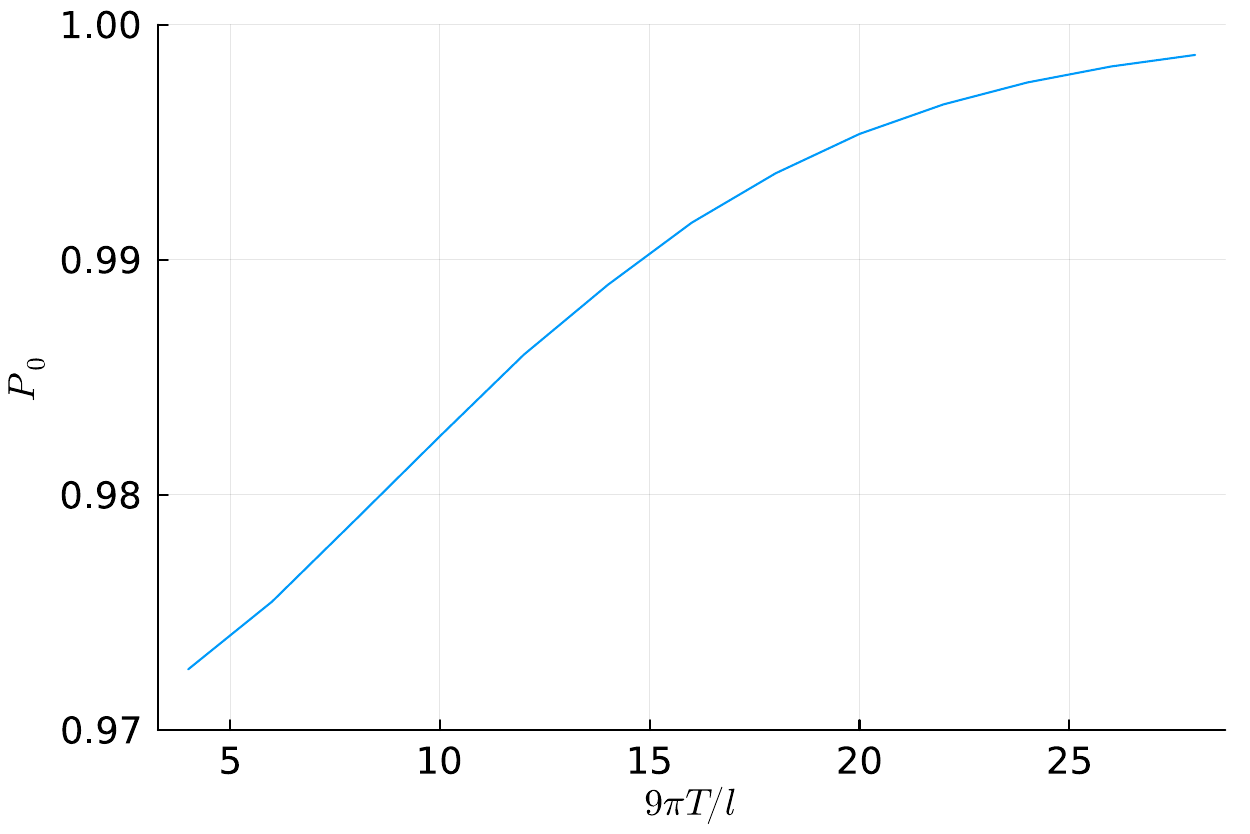}
    \caption{Probability of the $n=9$ mode being in the ground state of the associated thermal Hamiltonian as a function of $T$.}
    \label{fig:Notthegroundstate}
\end{figure}

Another way in which we can visualize the results of our simulation and connect it with the purity of the field mode being analyzed is the following: A Gaussian state can be seen as  a thermal state of some Hamiltonian; in particular, if the Gaussian state is pure, it can be interpreted as a ground state. Any measure of mixedness in a Gaussian state can therefore be equivalently understood in terms of the closeness of the state to the ground state of said Hamiltonian. Looking at the Hamiltonian for which a given Gaussian state is thermal, the probability of being in the ground state can be evaluated as a function of the symplectic eigenvalues of its covariance matrix~\cite{Brown_2013} as
\begin{equation}\label{eq:NottheGroundState}
    P_0 = \frac{2}{\nu + 1}.
\end{equation}
In Fig.~\ref{fig:Notthegroundstate}, we plot the probability of the $n=9$ mode being in the associated ground state\footnote{For the symplectic eigenvalue in Eq.~\eqref{eq:NottheGroundState}, we use the same time average as in Figs. \ref{fig:AverageSymplectic} and \ref{fig:VarianceSymplectic}.}.
We thus clearly see that as the cavity creation time $T$ increases, the mode in question approaches the ground state. In summary, our results thus point to the fact that the external potential is indeed capable of creating progressively purer localized modes of a quantum field, starting from the theory in free space.


One instance where these results can be particularly useful is in a protocol known as \emph{entanglement harvesting}. This consists of a process in which two independent localized quantum systems starting in a product state can become entangled after interacting with a quantum field. The final state of the probes after interacting with the field in this setup will generically contain contributions that can be divided into local terms (which affect the reduced state on each detector regardless of the existence of the other),  and nonlocal terms (which are responsible for the correlations between the two detectors). 

In rough intuitive terms, in order for the detectors to become entangled, the nonlocal correlations acquired by them should be strong enough to overcome the local terms experienced by each of them separately; in short, the nonlocal term must exceed the ``local noise'' on either detector. In this context, any initial mixedness in the state of the probes is detrimental to entanglement harvesting, as it effectively acts as another source of local noise that decreases the total amount of entanglement that can be potentially harvested. Therefore, in order for entanglement harvesting to be successful, it is important for the probes to be prepared in an initial state that is as pure as possible. This observation has been used to argue that particle detectors built out of modes of a relativistic quantum field have a fundamental limit to their ability to harvest entanglement at weak coupling~\cite{max}, since the high level of entanglement present in the QFT describing the probe system itself inevitably leads to some level of mixedness in the initial state of the detector. Our analysis indicates, however, that the mixedness of local modes of the probe field can be made systematically smaller, if one controls the dynamics external potential that provides the localization profile of the field. This reinforces the idea of using modes of localized quantum fields as particle detectors in a way that may not be fundamentally impeded to harvest entanglement~\cite{RuepReply, QFTPDHarvesting}.

\section{Conclusion}\label{sec:conclusions}

In this paper, we studied what happens when we confine degrees of freedom of a quantum field by dynamically growing cavity walls. In particular, our ultimate goal was the study of whether the confinement introduces inescapable mixedness in the localized degrees of freedom of the field in a way that could interfere with their use in relativistic quantum information protocols.

We showed that when we look at a confined mode with energy $E$ of a dynamically created cavity, if the cavity is created adiabatically enough (on a timescale $T \gtrsim E^{-1}$), the confined mode does not experience significant mixedness due to the raising of the cavity walls. Moreover, the localized modes do not get significantly excited by the potential growth. This provides a proof of principle that we can obtain localized modes of the field that can be made arbitrarily pure.

Overall, we believe that our analysis provides further evidence for the appropriate regimes where the adiabatic approximation is valid, and sheds light on some of the effects that appear away from that regime. Our results provide concrete computational tools to study the effect of time-dependent driving forces in QFT that complement previous analyses (see, e.g., ~\cite{Brown2015, fullhalfemptycavity, SmoothandSharp2018}). In particular, we improve over previous studies which used idealized sharp (even if time-dependent) boundary conditions by modeling the creation of cavities through time-dependent external potentials by also allowing for more realistic general shapes and smoother spatio-temporal dependence of the physical apparatus that effectively localizes the relevant degrees of freedom of the field.

One particularly important scenario where this result is relevant is the analysis of the entanglement harvesting protocol, where the mixedness of localized modes of quantum fields has been argued to be an unavoidable hindrance to one's ability to harvest entanglement with fully relativistic probes~\cite{max}. It is technically true that the Reeh-Schlieder theorem dictates that the reduced state of any local mode of a field theory in a finite region of space cannot be exactly pure unless there are regions of space where the confining potential diverges; however, our work shows that if the cavity walls are created `slowly enough'  the mixedness of localized modes of the field can be made arbitrarily negligible.\footnote{To get an idea of what “slow enough” means, we can point to the fact that the scale of adiabaticity for an optical cavity (trapping visible light) is roughly given by $T\sim 10^{-15}$ s.}

In future work, it would be interesting to study the structure of the entanglement between field modes under the effect of time-dependent potentials in more detail. We expect that the mixedness observed in the modes displayed in Fig.~\ref{fig:SymplecticEigenvalues} is mostly due to entanglement with field degrees of freedom outside of the region where the mode is supported. However, there may also be a non-trivial amount of internal entanglement among the modes within the same region of space where the cavity is being created. If this is true, then it would be possible to make an even better-informed choice of local modes via a local symplectic transformations that finds a new set of localized modes that are as uncorrelated as possible within the interior of the cavity for the specific potential that we are growing (along the lines of, e.g.,~\cite{kelly}).


Our main tool in this work consisted of a stable and convergent numerical method for evolving a scalar quantum field's two-point function in $(1+1)$-dimensions, given arbitrary initial data for the two-point function and its time derivatives at some initial time. This method then allowed us to reliably obtain the energy density of the field in a subregion of space and the covariance matrix of selected modes of the field. It must be noted, however, that these methods can be useful in a much broader set of problems. Promising avenues for future research extending these results include the study of the dynamical Casimir effect~\cite{DynamicalCasimir1, DynamicalCasimir2, DynamicalCasimir2other, wilson2011, DynamicalCasimir3, Velasco2022, Jean2025} with moving potentials instead of sharp reflecting mirrors. On another front, an extension of the methods outlined in Sec.~\ref{sec:numerics} to higher dimensions and general static background metrics in a memory-efficient way (through the techniques in e.g.~\cite{Zenger1991,SparseGrids}) is currently in progress. In the long term, we hope that this can provide a framework to address more advanced problems in quantum field theory in curved spacetimes that are outside the realm of capabilities for currently known methods.

\begin{acknowledgements}
    EMM acknowledges the support of the NSERC Discovery program as well as his Ontario Early Researcher Award. BSLT acknowledges financial support from the Mike and Ophelia Lazaridis Fellowship. BR acknowledges the Ontario Provincial Government for financial support from the Queen Elizabeth II Graduate Scholarship in Science \& Technology. Research at Perimeter Institute is supported in part by the Government of Canada through the Department of Innovation, Science and Industry Canada and by the Province of Ontario through the Ministry of Colleges and Universities. 
\end{acknowledgements}

\appendix

\section{CFL condition for waves with external potentials}\label{AppendixBoris}
When performing numerical simulations, it can be necessary to impose restrictions on the size of the time step to ensure that the numerical method in question is stable. One common method of finding these restrictions is known as Von Neumann Stability Analysis. This analysis relies on decomposing the perturbations in each grid point in terms of the Fourier modes with a certain wavelength. More explicitly, we write the perturbation as
\begin{equation}
    u^{n}_j = \hat{u}_n e^{ikx_j},
\end{equation}
where $\hat{u}$ is the amplitude of the wave at a given time, and $k$ is the wavenumber for the given perturbation. Since our grid spacing will be uniform, we take $x_j = j\Delta x$ and define $\theta \equiv k\Delta x$ as the ratio of the grid spacing and the wavelength. This provides a simplified form for our decomposition given by 
\begin{equation}
    u^{n}_j = \hat{u}_n e^{ij\theta}.
\end{equation}

Upon substituting this result into our FD method in Eq. \eqref{eq:DiscreteEvolutiont'}, we obtain 
\begin{align}
    \hat{u}_{n+1}e^{ij\theta} = 2&\hat{u}_ne^{ij\theta}-\hat{u}_{n-1}e^{ij\theta}\nonumber\\    &+C^2\left(\hat{u}_ne^{i(j+1)\theta}-2\hat{u}_ne^{ij\theta}+\hat{u}_ne^{i(j-1)\theta}\right)\nonumber\\
    &-2\Delta t^2V^n_i\hat{u}_ne^{ij\theta},\nonumber\\
\end{align}
which yields 
\begin{align}
    \hat{u}_{n+1} +\hat{u}_{n-1} =\hat{u}_n(2&+2C^2\cos(\theta)\nonumber\\
    &-2C^2-2\Delta t^2 V^n_i).\label{eq:stabilitytestUn}
\end{align}
A solution to this recursion relation can be found by assuming 
$\hat{u}_n = \lambda^n$. Using this ansatz in Eq.~\eqref{eq:stabilitytestUn}, we obtain a simple quadratic equation for $\lambda$,
\begin{equation}
    \lambda^2-M\lambda+1 =0,
\end{equation}
where 
\begin{equation}
    M = 2+2C^2\cos(\theta)-2C^2 -2\Delta t^2V^n_i.
\end{equation}
The roots of this equation are then given by
\begin{equation}\label{eq:quadraticeqlambda}
    \lambda_{\pm} = \dfrac{M}{2} \pm \dfrac{1}{2}\sqrt{M^2 - 4}.
\end{equation}
In order for the method to be stable (i.e., for the error to be bounded as $n\rightarrow \infty$), we must have $|\lambda_{\pm}|\leq 1$. If \mbox{$M^2>4$}, then at least one of the roots of Eq.~\eqref{eq:quadraticeqlambda} is greater than $1$ in magnitude: if $M>2$ then that root is $\lambda_+$, and if $M<-2$, it is $\lambda_-$. Therefore, the method is only stable if we have $M^2 <4$, in which case the two roots $\lambda_{\pm}$ are complex and given by 
\begin{equation}
    \lambda_{\pm} = \dfrac{M}{2} \pm \dfrac{\ii}{2}\sqrt{4 - M^2}.
\end{equation}
For $\lambda_{\pm}$ given by the expression above, with $4 - M^2 > 0$, we always have $\abs{\lambda_\pm} = 1$; i.e., both solutions are pure phases, and therefore satisfy the stability criteria. We thus simply need
\begin{equation}
    M^2\leq4 \iff -2\leq M\leq 2.
\end{equation}
In terms of $\Delta t$, $C$, and $V_i^n$, this translates to
\begin{equation}\label{eq:ineqM}
    -2\leq 2+2C^2\cos(\theta)-2C^2-2\Delta t^2 V^n_i\leq2.
\end{equation}
This equation must hold for all values of $\theta$ (i.e., for all wavevectors $k$) and at all points on the grid (i.e., for values of $i$). The second requirement is nontrivial because the potential $V_i^n$ is not uniform in space and time. 

The second inequality in~\eqref{eq:ineqM} is satisfied identically, since $\cos\theta - 1 \leq 0$ for all values of $\theta$ and $V_{i}^n\geq 0$ at all points on the grid. Therefore, the only inequality we need to check is the first one. Minimizing over $\theta$, we find that $\Delta t$ must obey
\begin{align}
    -2 &\leq 2 - 4C^2 - 2 \Delta t^2 V_i^n \nonumber \\
    \Rightarrow 4 &\geq 4 C^2 + 2\Delta t^2 V_i^n. \label{eq:ineqVin}
\end{align}
By now maximizing the right-hand side of the inequality in Eq.~\eqref{eq:ineqVin} and defining \hbox{$\Tilde{V} = \max_{\{n,i\}\in\mathcal{D}}V^n_i$}, we find 
\begin{align}
    4&\geq 4C^2 +2\Delta t^2\Tilde{V}\nonumber \\
    \implies \Delta t^2 &\leq \frac{2\Delta x^2}{2+\Tilde{V}\Delta x^2}, \label{eq:StabilityTime-Step}
\end{align}
where we used that $C$ is given explicitly by $C = \Delta t/\Delta x$.
Eq.~\eqref{eq:StabilityTime-Step} gives the relation between the choice of time step and spatial grid spacing necessary for for the numerical method to be stable. We must choose our time step in accordance with the above equation once we have selected a spatial grid spacing.

\twocolumngrid
\bibliography{references.bib}

\end{document}